\documentclass[journal,draftcls,onecolumn,12pt,twoside]{IEEEtran}

\normalsize

\ifCLASSINFOpdf
 
\else
  
\fi

\usepackage{cite}
\usepackage{enumerate}
\usepackage{amsmath,amssymb,amsfonts}
\usepackage{algorithmic}
\usepackage{graphicx}
\usepackage{textcomp}
\usepackage{xcolor}
\usepackage{float}
\usepackage{amsthm}
\usepackage{graphicx}
\usepackage{epstopdf}
\usepackage{amsmath,bm}
\usepackage{amsfonts}
\usepackage{amssymb}
\usepackage{color}
\usepackage{multirow}
\usepackage{multicol}
\usepackage{soul,xcolor}
\usepackage{subcaption}
\definecolor{orange}{RGB}{0,112,192}
\newtheorem{lemma}{Lemma}
\newtheorem{thm}{Theorem}
\newtheorem{remark}{Remark}
\DeclareMathOperator{\tr}{tr}
\DeclareMathOperator{\sinr}{SINR}
\DeclareMathOperator{\DS}{DS}
\DeclareMathOperator{\BU}{BU}
\DeclareMathOperator{\PC}{PC}
\DeclareMathOperator{\NI}{NI}
\DeclareMathOperator{\R}{R}
\begin{document}
\makeatletter
\newcommand*{\rom}[1]{\expandafter\@slowromancap\romannumeral #1@}
\makeatother
\title{Large-Scale Fading Precoding for Spatially Correlated Rician Fading with Phase Shifts
}

\author{\"Ozlem Tu\u{g}fe Demir,~\IEEEmembership{Member,~IEEE,} and
	Emil Bj\"ornson,~\IEEEmembership{Senior Member,~IEEE}
	\thanks{ A part of this paper was presented at ICASSP 2020\cite{conf}. The authors are with the Department of Electrical Engineering
		(ISY), Linköping University, 581 83 Linköping, Sweden (e-mail: ozlem.tugfe.demir@liu.se, emil.bjornson@liu.se)}
	\thanks{This work was partially supported by ELLIIT and the Wallenberg AI, Autonomous Systems and Software Program (WASP) funded by the Knut and Alice Wallenberg Foundation.}}

\maketitle

\begin{abstract}
We consider large-scale fading precoding (LSFP), which is a two-layer precoding scheme in the downlink of multi-cell massive MIMO (multiple-input multiple-output) systems to suppress inter-cell interference. We obtain the closed-form spectral efficiency (SE) with LSFP at the central network controller and maximum ratio precoding at the base stations (BSs) using the linear minimum mean-squared error or least squares channel estimators. The LSFP weights are designed based on the long-term channel statistics and two important performance metrics are optimized under the per-BS transmit power constraints. These metrics are sum SE and proportional fairness, where the resulting optimization problems are non-convex. Two efficient algorithms are developed to solve these problems by using the weighted minimum mean-squared error and the alternating direction method of multipliers methods. Moreover, two partial LSFP schemes are proposed to reduce the fronthaul signaling requirements. Simulations quantify the performance improvement of LSFP over standard single-layer precoding schemes and identify the specific advantage of each optimization problem.
\end{abstract}


\IEEEpeerreviewmaketitle

\section{Introduction}
\label{sec:intro}
Massive MIMO (multiple-input multiple-output) has received great attention and been analyzed from several theoretical and practical perspectives since the idea of deploying an infinite number of antennas at the base stations (BSs) first appeared in the seminal paper \cite{marzetta}. It has been seen as an essential technology for future wireless systems for the last decade due to its ability to spatially multiplex a large number of users on the same time and frequency resources \cite{erik_book,emil_book,ten_myths, utshick, caire2, caire1}. Although it has been well known that an antenna array can direct beams efficiently to the desired points and the directivity increases with the number of antennas for a long time in array signal processing, massive MIMO comes with not only using hundreds of antennas but also simple linear processing schemes by exploiting channel reciprocity in the uplink and downlink transmissions. Now, massive MIMO is one of the key components of 5G cellular systems and commercial deployments started in 2018 \cite{massive_mimo_reality}.

In the canonical form of massive MIMO, time division duplex (TDD) operation is adopted and channels between the BS and users are estimated in the uplink training phase of each coherence block and linear combiners and precoders for uplink data decoding and downlink data transmission, respectively, are selected based on the available channel estimates at a particular BS \cite{emil_book}. In \cite{marzetta}, it was shown that the effect of noise and intra-cell interference can be mitigated completely in the asymptotic region where the number of antennas at the BS goes to infinity. However, when there are multiple cells in the network and users in different cells share the same pilot sequence, the resulting non-orthogonal pilot transmission limits the spectral efficiency (SE), i.e., SE does not grow unboundedly with the number of BS antennas. This limitation is the result of interference from the other cells' users whose pilot signals are not orthogonal to the considered cell's users and this effect is called pilot contamination \cite{pilot_contamination}.

Several remedies have been proposed in the literature to mitigate the adverse effects of pilot contamination. One trivial remedy is to increase the length of the pilot signals to make them all orthogonal throughout the network. However, due to limited coherence block length in practical communication systems, this approach is not efficient \cite{chien_lsfd},\cite{interference_marzetta}. Assigning pilot sequences in a smart manner in a multi-cell network is another effort to suppress the pilot contamination effect \cite{pilot_scheduling,smart_pilot}. However, it may require solving a combinatorial optimization problem and not solve the problem at a desired level. Another solution to the pilot contamination problem is to exploit the spatial correlation among the BS antennas \cite{unlimited}. This method alleviates the performance upper bound that was observed for uncorrelated channels in \cite{marzetta} but, the multi-cell minimum mean-squared error (MMSE) decoding and precoding techniques from \cite{unlimited} have high computational complexity and require a large number of statistical parameters.

In \cite{pilot_precoding}, the authors showed that pilot contamination can be eliminated asymptotically when the BSs in different cells cooperate and a central network controller applies a second layer of decoding and precoding. Later these two-layer precoding and decoding techniques have been called large-scale fading precoding (LSFP) and decoding (LSFD). In \cite{interference_marzetta, uplink_interference,chien_lsfd,uplink_lsfd}, these approaches were elaborated. The prior work \cite{interference_marzetta} considered spatially uncorrelated Rayleigh fading channel model and the LSFP and LSFD weights are optimized using the max-min fairness criterion. Then, \cite{chien_lsfd} and \cite{uplink_lsfd} derived the SE with LSFD in the uplink for spatially correlated Rayleigh fading channels. Unlike other works, \cite{chien_lsfd} considered sum SE maximization objective, but only for uplink data transmission. 

We believe that there is an important gap regarding the analysis of LSFP using more realistic channel models and different optimization criteria since downlink operation was only considered for spatially uncorrelated Rayleigh fading and max-min fairness optimization in massive MIMO literature. To the best of authors' knowledge, this paper is the first work that considers the spatially correlated Rician fading with random phase shifts for the design of LSFP. The main contributions are:
\begin{itemize}
	
	\item We derive the downlink SE for a finite number of antennas with LSFP. In particular, we derive the closed-form SE expressions for both linear MMSE (LMMSE) and least squares (LS) channel estimate-based local maximum ratio (MR) precoding. 
	
	\item We consider sum SE and proportional fairness maximization problems with per-BS transmit power constraints, which are not considered before in the LSFP context. The optimization variables are LSFP weighting coefficients. To obtain a stationary point for the resulting non-convex problems, two efficient block coordinate descent algorithms are proposed by exploiting the weighted MMSE reformulation \cite{stationary} and implementing alternating direction method of multipliers (ADMM) \cite{boyd_admm}. The resulting algorithms have closed-form updates and use the structure of the problems to achieve much lower complexity than general-purpose numerical solvers.
	
	\item We propose two heuristic partial LSFP schemes to reduce the fronthaul signaling load between the central network controller and BSs. 
	
\end{itemize}
In the numerical results, we show that LSFP improves the SE of the users uniformly compared to the single-layer precoding with both cooperative and local power allocation. We provide several insights into the design of LSFP in a full or partial manner with different optimization criteria.
\section{System Model}

We consider a cellular network with $L$ cells. Each cell is composed of an $M$-antenna base station (BS) and $K$ single-antenna users. We assume  all BSs are connected to a central network controller in accordance with the existing literature \cite{chien_lsfd}, \cite{interference_marzetta}. The conventional block-fading model \cite{erik_book} is assumed where the channel between each BS antenna and user is a complex static scalar in one coherence block of $\tau_c$ channel uses and take independent realization in each block. In this paper, we assume TDD operation and, hence, channel reciprocity holds. We concentrate on the downlink part of the data transmission where the BSs serve the users in their cells with the aid of LSFP at the central network controller. Each coherence block is divided into two phases: uplink training and downlink data transmission. In the uplink training phase, all users send their assigned  pilot sequences of length $\tau_p$ and the BSs estimate the channel coefficients to design local precoding vectors that are matched to the estimated small-scale fading. The remaining $\tau_c-\tau_p$ samples are used for downlink data transmission. In accordance with the existing literature on massive MIMO, no downlink pilots are sent to the users and the users rely on channel statistics \cite{emil_book}.

Let ${\bf g}_{lk}^{r}\in \mathbb{C}^{M}$  denote the channel vector between user $k$ in cell $l$ and BS $r$.  We consider spatially correlated Rician fading channels, which is the first novelty of this paper in the context of LSFP. This means each channel realization can be expressed as
\begin{align} \label{eq:channel}
& {\bf g}_{lk}^{r}=e^{j\theta_{lk}^r}{{\bf \bar{g}}}_{lk}^{r}+{\bf \tilde{g}}_{lk}^{r},
\end{align}
where $e^{j\theta_{lk}^r}{\bf \bar{g}}_{lk}^{r}\in \mathbb{C}^{M}$  denotes the line-of-sight (LOS) component with some phase shift $\theta_{lk}^r$ common to all the antennas. The deterministic part of the LOS component, ${\bf \bar{g}}_{lk}^{r}$, is the array steering vector that is determined by the array geometry at BS $r$ and the angle of user $k$ in cell $l$ with respect to it. The other term of the channel, i.e., ${\bf \tilde{g}}_{lk}^{r}$, is the non-line-of-sight (NLOS) component and it is circularly symmetric Gaussian random vector with spatial covariance matrix ${\bf R}_{lk}^r \in \mathbb{C}^{M\times M}$, i.e., ${\bf \tilde{g}}_{lk}^{r} \sim \mathcal{N}_{\mathbb{C}}({\bf 0}_M,{\bf R}_{lk}^{r})$. Note that the vectors $\left\{{\bf \bar{g}}_{lk}^{r}\right\}$ and covariance matrices $\left\{{\bf R}_{lk}^{r}\right\}$ describe the long-term channel effects and they are fixed throughout the transmission. We assume that all BSs have the knowledge of $\left\{{\bf \bar{g}}_{lk}^{r}\right\}$ and covariance matrices $\left\{{\bf R}_{lk}^r\right\}$ in accordance with the multi-cell massive MIMO literature \cite{chien_lsfd}, \cite{ozge_massive}.\footnote{Please see \cite[Sec. 3.3.3]{emil_book} for the estimation of covariance matrices where more samples compared to small-scale channel estimation are available. In fact, the matrices can be estimated with relatively small amount of samples so that almost the same performance as the perfect statistical knowledge can be obtained. Similar methods can be used for the estimation of $\left\{{\bf \bar{g}}_{lk}^{r}\right\}$.} However compared to most of the existing literature, we consider a more realistic scenario in which phase of the LOS component varies at the same pace as the small-scale fading, due movement, and the phase is unknown. We assume the random phase shifts $\left\{\theta_{lk}^r\right\}$ are distributed uniformly on $[0,2\pi)$ \cite{ozge_cell_free}.

\subsection{Uplink Pilot Transmission and Channel Estimation}

All the cells use a common set of $\tau_p=K$ mutually orthogonal pilots where the pilots are distributed among the $K$ users in each cell in a disjoint manner.\footnote{Note that the best method in terms of channel estimation quality is to assign $LK$ mutually orthogonal pilots. However, the pilot resources are limited due to a fixed coherence block length, i.e., $\tau_p$ should be less than $\tau_c$. Even if $LK<\tau_c$, reserving a large portion of coherence block for uplink pilot transmission is not usually a good option due to the reduced downlink channel uses, and, hence, the SE.} Let $\bm{\varphi}_k \in \mathbb{C}^{\tau_p}$ denote the pilot sequence which is assigned to user $k$ in each cell where $\Vert\bm{\varphi}_{k}\Vert^2=\tau_p$ and $\bm{\varphi}_k^H\bm{\varphi}_{k^{\prime}}=0$, $\forall k^{\prime}\neq k$. Due to the pilot re-use between different cells, there is interference in the pilot transmission and so-called pilot contamination occurs.

During the uplink training phase, the received pilot signal ${\bf Z}_l\in \mathbb{C}^{M\times \tau_p}$ at BS $l$ is given by
\begin{align}\label{eq:pilot}
& {\bf Z}_{l}=\sum_{r=1}^L\sum_{k=1}^{K}\sqrt{\eta}{\bf g}_{rk}^{l}\bm{\varphi}_{k}^T+{\bf N}_l,
\end{align}
where $\eta$ is the pilot transmit power and the additive noise matrix ${\bf N}_{l}\in \mathbb{C}^{M \times \tau_p}$ has i.i.d. $\mathcal{N}_{\mathbb{C}}(0,\sigma^2)$ random variables. Then, the sufficient statistics for the channel information of user $k$ in cell $l$ is obtained as 
\begin{align}\label{eq:suff-stats} 
& {\bf z}_{lk}=\frac{{\bf Z}_l\bm{\varphi}_k^{*}}{\sqrt{\tau_p}}=\sqrt{\tau_p\eta}\sum_{r=1}^L {\bf g}_{rk}^l+{\bf \tilde{n}}_{lk},
\end{align}
where $\tilde{\bf n}_{lk}\triangleq{\bf N}_l\bm{\varphi}_k^{*}/\sqrt{\tau_p}$ has the distribution $\mathcal{N}_{\mathbb{C}}({\bf 0}_M,\sigma^2{\bf I}_M)$.

If the phase shifts $\{\theta_{lk}^r\}$ are not known, deriving an MMSE-based channel estimator is very hard since we do not have a linear Gaussian signal model. One possible approach is to estimate the channels using the LMMSE estimator that is the conventional benchmark in the massive MIMO literature and it is considered for the Rician fading channels with unknown phase shifts in \cite{ozge_cell_free} in the context of cell-free massive MIMO with single-antenna access points.\footnote{Note that due to the common phase shift affecting multiple antennas and, hence, correlation among the antennas, the channel estimation considered in this paper results different expressions from the single-antenna case in \cite{ozge_cell_free}.} The LMMSE estimation of the channel between user $k$ in cell $l$ and BS $l$ based on the sufficient statistics in \eqref{eq:suff-stats} is given by
\begin{align}\label{eq:phase-unaware-lmmse} 
&{\bf \hat{g}}_{lk}^l=\sqrt{\tau_p\eta}{\bf \overline{R}}_{lk}^l{\bf \Psi}_{lk}^{-1}{\bf z}_{lk},
\end{align}
where
\begin{align}
& {\bf \overline{R}}_{lk}^r\triangleq\mathbb{E}\left\{{\bf g}_{lk}^r\left({\bf g}_{lk}^r\right)^H\right\}= {\bf R}_{lk}^r+{\bf \bar{g}}_{lk}^r\left({\bf \bar{g}}_{lk}^r\right)^H, \label{eq:Rbar} \\
& {\bf \Psi}_{lk}\triangleq\mathbb{E}\left\{{\bf z}_{lk}{\bf z}_{lk}^H\right\}= \tau_p\eta\sum_{r=1}^L{\bf \overline{R}}_{rk}^l+\sigma^2{\bf I}_M \label{eq:Psi}.
\end{align}
The channel estimate ${\bf \hat{g}}_{lk}^l$ and the estimation error ${\bf e}_{lk}^l\triangleq{\bf g}_{lk}^l-{\bf \hat{g}}_{lk}^l$ are zero-mean uncorrelated random vectors with covariance matrices
\begin{align}
&\mathbb{E}\left\{{\bf \hat{g}}_{lk}^l\left({\bf \hat{g}}_{lk}^l\right)^H\right\}=\tau_p\eta{\bf \overline{R}}_{lk}^l{\bf \Psi}_{lk}^{-1}{\bf \overline{R}}_{lk}^l, \label{eq:covhath} \\
&\mathbb{E}\left\{{\bf e}_{lk}^l({\bf e}_{lk}^l)^H\right\}={\bf \overline{R}}_{lk}^l-\tau_p\eta{\bf \overline{R}}_{lk}^l{\bf \Psi}_{lk}^{-1}{\bf \overline{R}}_{lk}^l.
\end{align}

\begin{remark} Note that neither the channel estimate nor the estimation error are Gaussian for the LMMSE estimator. As a result, although they are uncorrelated, they are not independent. 
\end{remark}
\begin{remark} Note that the BSs do not need to estimate the channels of the users in the other cells in each coherence block. The central network controller has only the knowledge of all the channels' long-term channel statistics to design LSFP weights.
 \end{remark}
Note that the LMMSE-based channel estimator presented above requires the inversion of $M\times M$ matrices and becomes computationally demanding as the antenna number, $M$, increases. In the previous massive MIMO works that consider spatially correlated fading, a simpler estimation technique called \emph{element-wise MMSE (EW-MMSE)} which takes into account the diagonal components of the spatial covariance matrices $\{{\bf R}_{lk}^l\}$ is presented \cite{ozge_massive,chien_lsfd}. Inspired by this method, we can define the element-wise LMMSE (EW-LMMSE) estimate of the channel between user $k$ in cell $l$ and BS $l$ based on the sufficient statistics in \eqref{eq:suff-stats} and the considered Rician fading channels with the unknown phase shifts as follows:
\begin{align}\label{eq:phase-unaware-ew-lmmse} 
&{\bf \hat{g}}_{lk}^l=\sqrt{\tau_p\eta}{\bf \overline{D}}_{lk}^l{\bf \Lambda}_{lk}^{-1}{\bf z}_{lk},
\end{align}
where
\begin{align}
& {\bf \overline{D}}_{lk}^l\triangleq \text{diag}\left({\bf \overline{R}}_{lk}^l\right), \label{eq:Dbar} \\
& {\bf \Lambda}_{lk}\triangleq \text{diag}\left({\bf \Psi}_{lk}\right) \label{eq:Lambdabar}.
\end{align}
We note that the channel estimate ${\bf \hat{g}}_{lk}^l$ is just a scaled version of ${\bf z}_{lk}$ under the standard assumption that the correlation matrices $\left\{{\bf \overline{R}}_{lk}^r\right\}$ have equal diagonal elements \cite{chien_lsfd}. Similarly, the least squares (LS) estimator for the channel ${\bf g}_{lk}^l$ is also a scaled version of ${\bf z}_{lk}$. In the following part, we will use the MMSE and LS/EW-MMSE estimators to select the  maximum ratio (MR) local precoders. Note that for the LS/EW-MMSE-based channel estimation, we can use directly ${\bf z}_{lk}^{*}$ as MR local precoder since any scaling parameter can be included in the LSFP at the central network controller by power allocation.

\subsection{Downlink Data Transmission}
Let $s_{lk}$ denote the zero-mean unit-variance downlink symbol for user $k$ in cell $l$. In accordance with the previous work \cite{interference_marzetta}, we assume all the downlink symbols are accessible to the central network controller for LSFP.\footnote{In Section \rom{6}, we also consider partial LSFP where the central network controller has access only to a subset of the downlink symbols.} In the first step, the central network controller computes the symbols to be transmitted from each BS by conducting LSFP using the long-term statistics of the channel. Let $\left(a_{rk}^l\right)^{*}$ denote the complex weight applied to the data symbol of user $k$ in cell $r$ for the transmission of the combined signal $v_{lk}$ from the BS $l$, i.e.,
\begin{align}\label{eq:vlk} 
& v_{lk} =\sum_{r=1}^L(a_{rk}^{l})^{*}s_{rk}.
\end{align} 
Here, $v_{lk}$ is the precoded signal to be transmitted from BS $l$ for the users with index $k$. Hence, LSFP is applied to the data symbols of users sharing the same pilot sequence and each BS transmits a linear combination of pilot-sharing users' signals in an effort to precancel the pilot-contaminated interference that occurs between them.

In the second step, the central network controller sends the precoded symbols $\{v_{lk}\}_{k=1}^K$ to  BS $l$. Note that the power allocation is applied through the LSFP coefficients $\{a_{rk}^l\}$. Hence, there is no need for an additional localized power allocation at the BSs. In the last step, BS $l$ conducts local precoding based on the channel estimates obtained from the uplink pilot transmission.  Let ${\bf w}_{lk}^*$ denote the local precoding vector for the users that share the pilot sequence $k$. Then, the transmitted signal from the BS $l$ is 
\begin{align}\label{eq:xl} 
& {\bf x}_l =\sum_{k=1}^K{\bf w}_{lk}^{*}v_{lk}.
\end{align}

\section{Downlink Spectral Efficiency Analysis}
In this section, we will first derive a SE expression for the two-layer precoding (LSFP + local precoding) with any local precoding by utilizing the use-and-then-forget capacity bounding technique, which is standard in massive MIMO literature \cite{emil_book}. Then, we will compute a closed-form SE expression for the case of MR precoding.

The received signal at user $k$ in the cell $l$ is 
\begin{align} \label{eq:ylk}
& y_{lk}=\sum_{r=1}^L\left({\bf g}_{lk}^{r}\right)^T{\bf x}_{r}+ n_{lk},
\end{align}
where $n_{lk}$ is the additive Gaussian noise at the receiver of user $k$ in cell $l$ with zero-mean and variance $\sigma^2$. Let us rewrite \eqref{eq:ylk} as
\begin{align} \label{eq:ylk2} 
& y_{lk}=\DS_{lk}s_{lk}+\BU_{lk}s_{lk}+\sum_{\substack{r=1\\r\neq l}}^L\PC_{lk}^{r}s_{rk}+\sum_{r=1}^L\sum_{\substack{k^{\prime}=1\\ k^{\prime}\neq k}}^K\NI_{lk}^{rk^{\prime}}s_{rk^{\prime}}+n_{lk},
\end{align}
where $\DS_{lk}$, $\BU_{lk}$, $\PC_{lk}^{r}$, and $\NI_{lk}^{rk^{\prime}}$ represent the strength of the desired signal, the beamforming gain uncertainty, the pilot contamination, and the non-coherent interference which are defined as
\begin{align} 
&\DS_{lk}=\sum_{r=1}^L\left(a_{lk}^{r}\right)^{*}\mathbb{E}\left\{{\bf w}_{rk}^H{\bf g}_{lk}^{r}\right\} \label{eq:DS}, \\
&\BU_{lk}=\sum_{r=1}^L\left(a_{lk}^{r}\right)^{*}\left({\bf w}_{rk}^H{\bf g}_{lk}^{r}-\mathbb{E}\left\{{\bf w}_{rk}^H{\bf g}_{lk}^{r}\right\}\right) \label{eq:BU}, \\
& \PC_{lk}^{r}=\sum_{n=1}^L\left(a_{rk}^{n}\right)^{*}{\bf w}_{nk}^H{\bf g}_{lk}^{n}, \label{eq:PC} \\
&\NI_{lk}^{rk^{\prime}}= \sum_{n=1}^L\left(a_{rk^{\prime}}^{n}\right)^{*}{\bf w}_{nk^{\prime}}^H{\bf g}_{lk}^{n} \label{eq:NI}.
\end{align}
Note that the users do not know the actual value of the effective channel components ${\bf w}_{rk}^H{\bf g}_{lk}^{r}$ for the desired symbol, but only the expected value of the effective channel, which is $\DS_{lk}$ in \eqref{eq:DS}. The expected value is close to the effective channel in massive MIMO (thanks to channel hardening phenomenon \cite[Sec. 2.5]{emil_book}), if the precoding is selected based on the channel vector. Hence, the beamforming gain uncertainty resulting from imperfect channel state information in \eqref{eq:BU} is treated as interference. The pilot contamination from the users that use the same pilots in other cells and the non-coherent interference from the remaining users are the other sources of interference. The following lemma presents a lower bound on the downlink ergodic capacity by treating the signals other than desired signal as additive white Gaussian noise \cite[Sec. 4]{emil_book}.

\begin{lemma}  The downlink ergodic capacity of the user $k$ in cell $l$ for the given LSFP weights, is lower bounded by 
\begin{align}\label{eq:capacity}
\R_{lk}= \frac{\tau_c-\tau_p}{\tau_c}\log_2\left(1+\sinr_{lk}\right),
\end{align}
where $\sinr_{lk}$ is the effective SINR for user $k$ in cell $l$ and it is given by
\begin{align}\label{eq:sinr}
&\sinr_{lk}=\frac{\left|\DS_{lk}\right|^2}{\mathbb{E}\left\{\left|\BU_{lk}\right|^2\right\}+\sum\limits_{\substack{r=1 \\ r\neq l}}^L\mathbb{E}\left\{\left|\PC_{lk}^{r}\right|^2\right\}+\sum\limits_{\substack{r=1}}^L\sum\limits_{\substack{k^{\prime}=1 \\ k^{\prime}\neq k}}^K\mathbb{E}\left\{\left|\NI_{lk}^{rk^{\prime}}\right|^2\right\}+\sigma^2}.
\end{align}

\begin{IEEEproof}
	The lower bound expression in \eqref{eq:capacity} follows from \cite{erik_book} by noting that the interference  terms are mutually uncorrelated. 
\end{IEEEproof}
\end{lemma}
We will call $\R_{lk}$ in \eqref{eq:capacity} as the SE in the remainder of this paper. Let us define the following vectors and matrices for ease of notation in the following parts of the paper:
\begin{align}
& \bm{a}_{lk}\triangleq \left[ \ a_{lk}^1 \ \ldots \ a_{lk}^{L} \ \right]^T \in \mathbb{C}^{L},  \label{eq:alk}  \\
& \bm{b}_{lk}\triangleq \left[ \ b_{lk}^1 \ \ldots \ b_{lk}^{L}  \ \right]^T \in \mathbb{C}^{L}, \ \ \ b_{lk}^r\triangleq \mathbb{E}\left\{{\bf w}_{rk}^H{\bf g}_{lk}^{r}\right\} \label{eq:blk},  \\
&\bm{C}_{lkk^{\prime}}\in \mathbb{C}^{L \times L}, \ \ \ c_{lkk^{\prime}}^{rn}\triangleq \mathbb{E}\left\{{\bf w}_{rk^{\prime}}^H{\bf g}_{lk}^{r}\left({\bf g}_{lk}^{n}\right)^H{\bf w}_{nk^{\prime}}\right\} \label{eq:Clk},
\end{align}
where $c_{lkk^{\prime}}^{rn}=\left[\bm{C}_{lkk^{\prime}}\right]_{rn}$ is the $(r,n)$th element of the matrix $\bm{C}_{lkk^{\prime}}$.
Using the above definitions, the terms in the SINR expression in \eqref{eq:sinr} can be expressed as
\begin{align} 
&\left|\DS_{lk}\right|^2=\left|\bm{a}_{lk}^H\bm{b}_{lk}\right|^2, \label{eq:DS2} \\
&\mathbb{E}\left\{\left|\BU_{lk}\right|^2\right\}=\bm{a}_{lk}^H\bm{C}_{lkk}\bm{a}_{lk}-\left|\bm{a}_{lk}^H\bm{b}_{lk}\right|^2, \label{eq:BU2} \\
& \mathbb{E}\left\{\left|\PC_{lk}^{r}\right|^2\right\}=\bm{a}_{rk}^H\bm{C}_{lkk}\bm{a}_{rk}, \label{eq:PC2} \\
&\mathbb{E}\left\{\left|\NI_{lk}^{rk^{\prime}}\right|^2\right\}=\bm{a}_{rk^{\prime}}^H\bm{C}_{lkk^{\prime}}\bm{a}_{rk^{\prime}} \label{eq:NI2}.
\end{align}
Using the results in (\ref{eq:DS2})-(\ref{eq:NI2}), the SINR for user $k$ in cell $l$ in \eqref{eq:sinr} is written as
\begin{align}\label{eq:SE1}
&\sinr_{lk}=\frac{\left|\bm{a}_{lk}^H\bm{b}_{lk}\right|^2}{\sum\limits_{r=1}^L\bm{a}_{rk}^H\bm{C}_{lkk}\bm{a}_{rk}-\left|\bm{a}_{lk}^H\bm{b}_{lk}\right|^2+\sum\limits_{\substack{r=1}}^L\sum\limits_{\substack{k^{\prime}=1 \\ k^{\prime}\neq k}}^K\bm{a}_{rk^{\prime}}^H\bm{C}_{lkk^{\prime}}\bm{a}_{rk^{\prime}}+\sigma^2}.
\end{align}
 Note that the LSFD weights in the uplink can be selected independently for each user by maximizing a generalized Rayleigh quotient, as shown in \cite{chien_lsfd}. However, for the LSFP, the precoding weight vector $\bm{a}_{lk}$ affects not only the $\sinr_{lk}$ but also the interference level of all other users. Hence, a joint design is needed for the LSFP in the downlink.
 
Before optimizing the LSFP vectors, we will derive the closed-form SE expressions for MR local precoding by evaluating the expectations in (\ref{eq:blk})-(\ref{eq:Clk}). The following theorems present the SE for the MR local precoding vectors ${\bf w}_{lk}^{*}=\left({\bf \hat{g}}_{lk}^l\right)^{*}$ based on LMMSE estimate in \eqref{eq:phase-unaware-lmmse} and ${\bf w}_{lk}^{*}={\bf z}_{lk}^{*}$ based on scaled LS estimate in \eqref{eq:suff-stats}. 

\begin{thm}\label{the1} For a given set of LSFP coefficients, the SE of user $k$ in cell $l$ for the local precoding vectors $\left\{{\bf w}_{lk}^{*}=\left({\bf \hat{g}}_{lk}^l\right)^{*}\right\}$ (LMMSE estimates in \eqref{eq:phase-unaware-lmmse}) is given in \eqref{eq:capacity} with $\sinr_{lk}$ as in \eqref{eq:SE1} where the elements of $\bm{b}_{lk}$ and $\bm{C}_{lkk^{\prime}}$ are given by
	\begin{align}
	b_{lk}^{r}=&\mathbb{E}\left\{\left({\bf \hat{g}}_{rk}^r\right)^H{\bf g}_{lk}^{r}\right\}=\tau_p\eta\tr\left({\bf \Psi}_{rk}^{-1}{\bf \overline{R}}_{rk}^{r}{\bf \overline{R}}_{lk}^{r}\right), \label{eq:blkr1} \\
	c_{lkk}^{rr}=&\mathbb{E}\left\{\left({\bf \hat{g}}_{rk}^r\right)^H{\bf g}_{lk}^{r}\left({\bf g}_{lk}^{r}\right)^H{\bf \hat{g}}_{rk}^r\right\}\nonumber\\
	=&\tau_p^2\eta^2\left|\tr\left({\bf R}_{lk}^{r}{\bf \overline{R}}_{rk}^{r}{\bf \Psi}_{rk}^{-1}\right)\right|^2+2\tau_p^2\eta^2\Re\left\{\left({\bf \bar{g}}_{lk}^r\right)^H{\bf \overline{R}}_{rk}^r{\bf \Psi}_{rk}^{-1}{\bf \bar{g}}_{lk}^r\tr\left({\bf \Psi}_{rk}^{-1}{\bf \overline{R}}_{rk}^r{\bf R}_{lk}^r\right)\right\} \nonumber \\
	& +\tau_p\eta\tr\left({\bf \overline{R}}_{rk}^r{\bf \Psi}_{rk}^{-1}{\bf \overline{R}}_{rk}^r{\bf \overline{R}}_{lk}^r\right),  \label{eq:clkk-rr1} \\
	c_{lkk}^{rn}=&\mathbb{E}\left\{\left({\bf \hat{g}}_{rk}^r\right)^H{\bf g}_{lk}^{r}\left({\bf g}_{lk}^{n}\right)^H{\bf \hat{g}}_{nk}^n\right\}=b_{lk}^r\left(b_{lk}^n\right)^{*}, \ \ \ r\neq n, \label{eq:clkk-rn1} \\
	c_{lkk^{\prime}}^{rr}=&\mathbb{E}\left\{\left({\bf \hat{g}}_{rk^{\prime}}^r\right)^H{\bf g}_{lk}^{r}\left({\bf g}_{lk}^{r}\right)^H{\bf \hat{g}}_{rk^{\prime}}^r\right\}=\tau_p\eta\tr\left({\bf \overline{R}}_{rk^{\prime}}^r{\bf \Psi}_{rk^{\prime}}^{-1}{\bf \overline{R}}_{rk^{\prime}}^r{\bf \overline{R}}_{lk}^r\right), \ \ \  k^{\prime} \neq k, \label{eq:clkk2-rr1} \\
	c_{lkk^{\prime}}^{rn}=&\mathbb{E}\left\{\left({\bf \hat{g}}_{rk^{\prime}}^r\right)^H{\bf g}_{lk}^{r}\left({\bf g}_{lk}^{n}\right)^H{\bf \hat{g}}_{nk^{\prime}}^n\right\}=0, \ \ \ k^{\prime} \neq k, \ r \neq n. \label{eq:clkk2-rn1}
	\end{align}
	
	\begin{IEEEproof} Please see the Appendix \ref{the1_proof} for the proof.
	\end{IEEEproof}
\end{thm}
Note that using the LMMSE estimation-based local precoding ${\bf w}_{lk}^{*}=\left({\bf \hat{g}}_{lk}^l\right)^{*}$ requires the inversion of $M \times M$ matrices for both channel  estimation and the calculation of the closed-form SE as shown in Theorem~\ref{the1}. This becomes computationally demanding as the antenna number, $M$, increases. One computationally efficient approach for the BSs is to use ${\bf z}_{lk}^{*}$ as the local-precoder for the combination of the signals of the users with index $k$. Note that ${\bf z}_{lk}$ is the scaled version of LS channel estimate for these users. If the diagonal elements of the correlation matrices of the channels are assumed to be the same,  $\left\{{\bf z}_{lk}\right\}$ are also the scaled version of EW-LMMSE channel estimates.

\begin{thm}\label{the2} For a given set of LSFP coefficients, the SE of user $k$ in cell $l$ for the local precoding vectors $\left\{{\bf w}_{lk}^*={\bf z}_{lk}^{*}\right\}$ is given in \eqref{eq:capacity} with $\sinr_{lk}$ as in \eqref{eq:SE1} where the elements of $\bm{b}_{lk}$ and $\bm{C}_{lkk^{\prime}}$ are given by
\begin{align}
b_{lk}^{r}=&\mathbb{E}\left\{{\bf z}_{rk}^H{\bf g}_{lk}^{r}\right\}=\sqrt{\tau_p\eta}\tr\left({\bf \overline{R}}_{lk}^{r}\right), \label{eq:blkr2}\\
c_{lkk}^{rr}=&\mathbb{E}\left\{{\bf z}_{rk}^H{\bf g}_{lk}^{r}\left({\bf g}_{lk}^{r}\right)^H{\bf z}_{rk}\right\}=\tau_p\eta\left(\tr\left({\bf R}_{lk}^{r}\right)\right)^2+2\tau_p\eta\left({\bf \bar{g}}_{lk}^r\right)^H{\bf \bar{g}}_{lk}^r\tr\left({\bf R}_{lk}^{r}\right)+\tr\left({\bf \Psi}_{rk}{\bf \overline{R}}_{lk}^{r}\right) \label{eq:clkk-rr2}, \\
c_{lkk}^{rn}=&\mathbb{E}\left\{{\bf z}_{rk}^H{\bf g}_{lk}^{r}\left({\bf g}_{lk}^{n}\right)^H{\bf z}_{nk}\right\}=b_{lk}^r\left(b_{lk}^n\right)^{*}, \ \ \ r\neq n, \label{eq:clkk-rn2} \\
c_{lkk^{\prime}}^{rr}=&\mathbb{E}\left\{{\bf z}_{rk^{\prime}}^H{\bf g}_{lk}^{r}\left({\bf g}_{lk}^{r}\right)^H{\bf z}_{rk^{\prime}}\right\}=\tr\left({\bf \Psi}_{rk^{\prime}}{\bf \overline{R}}_{lk}^{r}\right), \ \ \  k^{\prime} \neq k, \label{eq:clkk2-rr2}\\
 c_{lkk^{\prime}}^{rn}=&\mathbb{E}\left\{{\bf z}_{rk^{\prime}}^H\right\}\mathbb{E}\left\{{\bf g}_{lk}^{r}\right\}\mathbb{E}\left\{\left({\bf g}_{lk}^{n}\right)^H\right\}\mathbb{E}\left\{{\bf z}_{nk^{\prime}}\right\}=0, \ \ \ k^{\prime} \neq k, \ r \neq n. \label{eq:clkk2-rn2}
\end{align}
\begin{IEEEproof} Please see the Appendix \ref{the2_proof} for the proof.
\end{IEEEproof}
\end{thm}

Note that the results obtained so far assume the local precoding vectors $\{{\bf w}_{lk}^*\}$ are not normalized. This does not pose any issue since the LSFP weights take the role of scaling to satisfy the per-BS transmit power constraints. As we show in the next part, we take the norms of the local precoding vectors into account in constructing the power constraints.

\section{LSFP Optimization for Sum SE Maximization}
\label{sec:sumSE}
In this section, we design the LSFP weights $\{a_{rk}^l\}$ in order to maximize the sum of all users' SEs under individual BS transmit power constraints. 

Let us define $\omega_{lk}\triangleq\mathbb{E}\left\{\Vert{\bf w}_{lk}\Vert^2\right\}$ that is given for different local precoding selections as
\begin{align}
&\omega_{lk}=\begin{cases} \tau_p\eta\tr\left({\bf \overline{R}}_{lk}^l{\bf \Psi}_{lk}^{-1}{\bf \overline{R}}_{lk}^l\right), &   \text{if } {\bf w}_{lk}={\bf \hat{g}}_{lk}^l \text{ (LMMSE) } \\
\tr\left({\bf \Psi}_{lk}\right),& \text{if } {\bf w}_{lk}={\bf z}_{lk}  \text{ (LS) }\end{cases}.
\end{align}
Then, the long-term transmit power of BS $l$ is given by
\begin{align} \label{eq:power}
P_{l}=&\mathbb{E}\left\{\Vert{\bf x}_l\Vert^2\right\}=\sum_{k=1}^K\omega_{lk}\sum_{r=1}^L|a_{rk}^l|^2.
\end{align}
 The sum SE maximization problem in terms of LSFP weight vectors $\left\{\bm{a}_{lk}\right\}$ can be expressed as 
\begin{align}
&\underset{\left\{\bm{a}_{lk}\right\}}{\text{maximize}} \ \ \ \sum_{l=1}^L\sum_{k=1}^K\log_2\left( 1+\sinr_{lk}\right) \label{eq:obj} \\
&\text{subject to} \ \ \ \sum\limits_{k=1}^K\omega_{lk}\sum\limits_{r=1}^L|a_{rk}^l|^2\leq\rho_d, \ \ l=1,\ldots,L,  \label{eq:const1} 
\end{align}
where $\rho_d$ is the maximum downlink transmission power of each BS and constant pre-log factor of $\R_{lk}$ in \eqref{eq:capacity} is neglected in the objective.

Note that the problem in (\ref{eq:obj})-(\ref{eq:const1}) is not convex and it is hard to obtain the global optimum solution with a reasonable complexity. To obtain an effective solution to this non-convex problem, we will use the weighted MMSE method to develop an efficient iterative algorithm as in \cite{chien_lsfd}. However, due to the different structure of the downlink LSFP than the uplink LSFD in \cite{chien_lsfd}, the steps of the algorithms are different. Moreover, we utilize the special structure of our problem to develop a low-complexity ADMM algorithm to solve the sub-problems of the iterative algorithms optimally.

To express the optimization problem in (\ref{eq:obj})-(\ref{eq:const1}) as a weighted MMSE problem, we first recall the received downlink signal at user $k$ in cell $l$ from \eqref{eq:ylk2}. The desired signal $s_{lk}$ at user $k$ in cell $l$ is decoded by applying the receiver weight $u_{lk}^*\in\mathbb{C}$, i.e., $\hat{s}_{lk}=u_{lk}^*y_{lk}$. Then, the mean-square error (MSE) is given by
\begin{align}\label{eq:elk}
e_{lk}=\mathbb{E}\left\{|\hat{s}_{lk}-s_{lk}|^2\right\}=|u_{lk}|^2\Bigg(\sum_{r=1}^L\sum_{k^{\prime}=1}^K\bm{a}_{rk^{\prime}}^H\bm{C}_{lkk^{\prime}}\bm{a}_{rk^{\prime}}+\sigma^2\Bigg)-2\Re\left\{u_{lk}^*\bm{a}_{lk}^H\bm{b}_{lk}\right\}+1.
\end{align}
Note that $e_{lk}$ is a convex function of the beamformer weight $u_{lk}$ and the optimum $u_{lk}$ is obtained as
\begin{align} \label{eq:ulk}
u_{lk}=\frac{\bm{a}_{lk}^H\bm{b}_{lk}}{\sum_{r=1}^L\sum_{k^{\prime}=1}^K\bm{a}_{rk^{\prime}}^H\bm{C}_{lkk^{\prime}}\bm{a}_{rk^{\prime}}+\sigma^2},
\end{align} 
where the optimum $e_{lk}$ can be shown to be equal to $e_{lk}=1/\left(1+\sinr_{lk}\right)$. After introducing the weights $d_{lk}\geq 0$ for the MSE $e_{lk}$, we formulate the following weighted MMSE problem 
\begin{align}
&\underset{\left\{\bm{a}_{lk},  \ u_{lk}, \ d_{lk}\geq 0 \right\}}{\text{minimize}} \ \ \ \sum_{l=1}^L\sum_{k=1}^K\Big(d_{lk}e_{lk}-\ln\left( d_{lk}\right)\Big) \label{eq:obj2} \\
&\hspace{0.8cm}\text{subject to} \ \ \ \sum\limits_{k=1}^K\omega_{lk}\sum\limits_{r=1}^L|a_{rk}^l|^2\leq\rho_d, \ \ l=1,\ldots,L,  \label{eq:const12} 
\end{align}
where $e_{lk}$ is as in \eqref{eq:elk}. The weighted MMSE problem in (\ref{eq:obj2})-(\ref{eq:const12}) is equivalent to the original sum SE maximization problem  (\ref{eq:obj})-(\ref{eq:const1}) in the sense that they have the same global optimum solution. The equivalency of two problems easily follows from the fact that the optimum $d_{lk}$ for the above problem is $1/e_{lk}$, that is equal to $1+\sinr_{lk}$. Hence, we obtain the sum SE maximization problem with the same constraints.

The following theorem states that applying block coordinate descent to the problem  (\ref{eq:obj2})-(\ref{eq:const12}),  i.e., alternately minimizing it by keeping the other variables as constant generates a stationary point to the problems (\ref{eq:obj})-(\ref{eq:const1}) and (\ref{eq:obj2})-(\ref{eq:const12}).

\begin{thm} \label{the3}By iteratively solving (\ref{eq:obj2})-(\ref{eq:const12}) in an alternating manner for the blocks of variables $\left\{u_{lk}\right\}$, $\left\{d_{lk}\right\}$, $\left\{\bm{a}_{lk}\right\}$, the variables will converge to a stationary point of (\ref{eq:obj2})-(\ref{eq:const12}). Furthermore, $\left\{\bm{a}_{lk}\right\}$ converge to a stationary point of  (\ref{eq:obj})-(\ref{eq:const1}).
	
	\begin{IEEEproof} We first note that the problem in (\ref{eq:obj2})-(\ref{eq:const12}) is strictly convex in the blocks of variables $\left\{u_{lk}\right\}$, $\left\{d_{lk}\right\}$, $\left\{\bm{a}_{lk}\right\}$. Hence, the objective function is improved at each iteration. Furthermore, the objective function is lower bounded. So, block coordinate descent algorithm applied to this problem has a limit point. Since the constraints are separable for all the blocks of variables, the theorem can be proven similar to  \cite[Theorem 3]{stationary}.
		\end{IEEEproof}
\end{thm}

In the block coordinate descent algorithm we propose, the optimization problem in (\ref{eq:obj2})-(\ref{eq:const12}) is solved by treating only one of the blocks of variables $\left\{u_{lk}\right\}$, $\left\{d_{lk}\right\}$, $\left\{\bm{a}_{lk}\right\}$ as variable and keeping the others fixed at the previously obtained values. The update for the block of variables $\left\{u_{lk}\right\}$ that optimizes (\ref{eq:obj2})-(\ref{eq:const12}) for fixed $\left\{d_{lk}\right\}$, $\left\{\bm{a}_{lk}\right\}$ is already given in \eqref{eq:ulk}. The optimum $\left\{d_{lk}\right\}$ by keeping the other variables as constant can be obtained as $d_{lk}=1/e_{lk}$. Let us consider the optimization problem (\ref{eq:obj2})-(\ref{eq:const12}) in terms of $\left\{\bm{a}_{lk}\right\}$ for fixed $\left\{u_{lk}\right\}$, $\left\{d_{lk}\right\}$, which is a quadratically-constrained quadratic program (QCQP) and strongly convex. However, the closed-form solution cannot be obtained due to more than one constraint. We can solve this problem with general-purpose numerical solvers, such as CVX, but these would not exploit the specific structure of the problem and become time-consuming as the problem size increases. Recently, several efficient consensus ADMM-based methods have been developed for both convex and non-convex QCQP problems, which are shown to require much less time in comparison to the general-purpose numerical solvers \cite{consensus_admm, admm_fast}. In this paper, we propose an ADMM-based algorithm to solve the QCQP problem (\ref{eq:obj2})-(\ref{eq:const12}) for  $\left\{\bm{a}_{lk}\right\}$, which converges to the optimum solution since the problem is convex \cite{boyd_admm}.

First, we introduce the variables 
\begin{align} \label{eq:alk-tilde}
&\bm{\tilde{a}}_{lk}=\bm{\Omega}_{k}\bm{a}_{lk}, \quad l=1,\ldots,L, \quad k=1,\ldots,K,
\end{align}
where $\bm{\Omega}_{k}\in\mathbb{R}^{L\times L}$ denotes the diagonal matrix whose $r$th diagonal element is $\sqrt{\omega_{rk}}$. Hence, the $r$th element of $\bm{\tilde{a}}_{lk}$ is $\sqrt{\omega_{rk}}a_{lk}^r$. To obtain an efficient ADMM-based algorithm with closed-form updates, we reformulate the convex problem in (\ref{eq:obj2})-(\ref{eq:const12}) as
\begin{align}
&\underset{\left\{\bm{\tilde{a}}_{lk}, \ \bm{\bar{a}}_{lk}   \right\}}{\text{minimize}} \ \ \ \sum_{l=1}^L\sum_{k=1}^K\left( \bm{\tilde{a}}_{lk}^H\bm{F}_{lk}\bm{\tilde{a}}_{lk}-2\Re\left\{\bm{\tilde{a}}_{lk}^H\bm{f}_{lk}\right\} \right) \label{eq:obj2b} \\
&\hspace{0.1cm}\text{subject to} \ \ \ \sum\limits_{k=1}^K\sum\limits_{r=1}^L|\bar{a}_{rk}^l|^2\leq\rho_d, \ \ l=1,\ldots,L,  \label{eq:const12b} \\
&\hspace{2.3cm} \bm{\bar{a}}_{lk}=\bm{\tilde{a}}_{lk}, \quad l=1,\ldots,L, \quad k=1,\ldots,K, \label{eq:const22b} 
\end{align} 
where we have defined
\begin{align}
\bm{F}_{lk}&\triangleq \sum_{r=1}^L\sum_{k^{\prime}=1}^Kd_{rk^{\prime}}|u_{rk^{\prime}}|^2\bm{\Omega}_{k}^{-1}\bm{C}_{rk^{\prime}k}\bm{\Omega}_{k}^{-1}, \\
\bm{f}_{lk}&\triangleq d_{lk}u_{lk}^*\bm{\Omega}_{k}^{-1}\bm{b}_{lk}, \quad l=1,\ldots,L, \quad k=1,\ldots,K
\end{align}
for ease of notation. In \eqref{eq:const22b}, we have introduced a local copy of $\bm{\tilde{a}}_{lk}$ to obtain the closed-form updates in the ADMM-based algorithm by splitting the objective and the constraints. This technique is known as consensus ADMM \cite{boyd_admm}. Introducing the scaled dual variables $\left\{\bm{\hat{a}}_{lk}\right\}$ corresponding to the equality constraints in \eqref{eq:const22b} and the penalty parameter $\rho>0$ used in the augmented Lagrangian \cite{boyd_admm}, the steps of the ADMM algorithm for the problem (\ref{eq:obj2b})-(\ref{eq:const22b}) in scaled-form \cite{boyd_admm,admm_fast} are given by
\begin{enumerate}
\item Update the first block of primal variables, $\left\{\bm{\tilde{a}}_{lk}\right\}$, as
\begin{align}
&\left\{\bm{\tilde{a}}_{lk}\right\}  \leftarrow  \arg  \min_{\left\{\bm{\tilde{a}}_{lk}\right\}} \ \sum_{l=1}^L\sum_{k=1}^K\left( \bm{\tilde{a}}_{lk}^H\bm{F}_{lk}\bm{\tilde{a}}_{lk}-2\Re\left\{\bm{\tilde{a}}_{lk}^H\bm{f}_{lk}\right\}+\rho\Vert\bm{\bar{a}}_{lk}-\bm{\tilde{a}}_{lk}+\bm{\hat{a}}_{lk}\Vert^2 \right) \label{eq:step1}.
\end{align}
\item Update the second block of primal variables, $\left\{\bm{\bar{a}}_{lk}\right\}$, as
\begin{align}
& \left\{\bm{\bar{a}}_{lk}\right\}  \leftarrow  \arg  \min_{\left\{\bm{\bar{a}}_{lk}\right\}} \ \rho\sum_{l=1}^L\sum_{k=1}^K\Vert\bm{\bar{a}}_{lk}-\bm{\tilde{a}}_{lk}+\bm{\hat{a}}_{lk}\Vert^2  \label{eq:step2-a}\\ 
&\hspace{1.8cm}\text{subject to} \ \ \sum\limits_{k=1}^K\sum\limits_{r=1}^L|\bar{a}_{rk}^l|^2\leq\rho_d, \ \ l=1,\ldots,L. \label{eq:step2-b}
\end{align}
\item Update the dual variables, $\left\{\bm{\hat{a}}_{lk}\right\}$, as
 \begin{align} \label{eq:dual-update}
& \bm{\hat{a}}_{lk}\leftarrow \bm{\bar{a}}_{lk}-\bm{\tilde{a}}_{lk}+\bm{\hat{a}}_{lk}, \quad l=1,\ldots,L, \quad k=1,\ldots,K.
 \end{align}
\end{enumerate}
Note that the dual variable updates in \eqref{eq:dual-update} are simple additions. The closed-form primal variable updates can be obtained by solving the unconstrained convex quadratic programming in \eqref{eq:step1} and the $L$ independent Euclidean projection problems in (\ref{eq:step2-a})-(\ref{eq:step2-b}) \cite{admm_fast}\footnote{The motivation for introducing the variables $\bm{\tilde{a}}_{lk}$ in \eqref{eq:alk-tilde} can now be understood well. With these variables, we obtain a closed-form solution to the Euclidean projection problem (\ref{eq:step2-a})-(\ref{eq:step2-b}), which would not possible due to the varying weights $\omega_{lk}$ in the original formulation.} are given by
\begin{align}
& \bm{\tilde{a}}_{lk}  \leftarrow \left(\bm{F}_{lk}+\rho{\bf I}_L\right)^{-1}\left(\bm{f}_{lk}+\rho\bm{\bar{a}}_{lk}+\rho\bm{\hat{a}}_{lk}\right), \quad l=1,\ldots,L, \quad k=1,\ldots,K, \label{eq:update-primal1} \\
& \bar{a}_{rk}^l \leftarrow \min\left\{\sqrt{\frac{\rho_d}{\sum_{r^{\prime}=1}^L\sum_{k^{\prime}=1}^K\left\vert\tilde{a}_{r^{\prime}k^{\prime}}^l-\hat{a}_{r^{\prime}k^{\prime}}^l\right\vert^2}}, \ 1\right\}\left(\tilde{a}_{rk}^l-\hat{a}_{rk}^l\right), \nonumber\\
&\hspace{2cm} \quad r=1\ldots,L, \quad k=1,\ldots,K, \quad l=1,\ldots,L. \label{eq:update-primal2}
\end{align}
We present the steps of the overall block coordinate descent algorithm with ADMM in Algorithm~1 where $(i)$ and $(j)$ denote the $i$th outer iteration of the block descent and the $j$th inner iteration of ADMM, respectively. 
\vspace{-0.7cm}
\begin{center}
	\linethickness{0.45mm}
	\line(1,0){485}
\end{center}
\vspace{-0.4cm}
{\bf Algorithm 1:} Block Coordinate Descent Algorithm with ADMM for Sum SE Maximization
\vspace{-1.2cm}
\begin{center}
	\linethickness{0.15mm}
	\line(1,0){485}
\end{center}
\vspace{-0.6cm}
{\bf 1)}  Initiate the outer iteration number: $i=0$. Choose the elements of  $\left\{\bm{a}_{lk}^{(0)}\right\}$ such that they are identical positive numbers that satisfy the per-BS power constraints in \eqref{eq:const1} with equality. \\
{\bf 2)}  Set $i \leftarrow i+1$ and update $\left\{u_{lk}^{(i)}\right\}$ using \eqref{eq:ulk} with $\left\{\bm{a}_{lk}^{(i-1)}\right\}$. \\
{\bf 3)} Update $\left\{d_{lk}^{(i)}\right\}$ as $d_{lk}^{(i)}=1/e_{lk}^{(i-1)}$ where $e_{lk}^{(i-1)}$ is evaluated in \eqref{eq:elk} using  $\left\{u_{lk}^{(i)}\right\}$ and $\left\{\bm{a}_{lk}^{(i-1)}\right\}$.  \\
{\bf 4)} Update $\left\{\bm{a}_{lk}^{(i)}\right\}$ as the solution of the following ADMM algorithm: \\
\indent {\bf 4a)} Initiate the inner iteration number: $j=0$. Initialize the second block of primal variables, $\bm{\bar{a}}^{(0)}$, randomly. Set the dual variables to zero: $\bm{\hat{a}}_{lk}^{(0)}={\bf 0}_L$. \\
\indent  {\bf 4b)} Set $j \leftarrow j+1$ and update $\bm{\tilde{a}}_{lk}^{(j)}$ using \eqref{eq:update-primal1} with $\bm{\bar{a}}_{lk}^{(j-1)}$ and $\bm{\hat{a}}_{lk}^{(j-1)}$.\\
\indent  {\bf 4c)} Update $\bm{\bar{a}}_{lk}^{(j)}$ using \eqref{eq:update-primal2} with $\bm{\tilde{a}}_{lk}^{(j)}$ and $\bm{\hat{a}}_{lk}^{(j-1)}$.\\
\indent  {\bf 4d)} Update $\bm{\hat{a}}_{lk}^{(j)}$ using \eqref{eq:dual-update} with $\bm{\tilde{a}}_{lk}^{(j)}$, $\bm{\bar{a}}_{lk}^{(j)}$, and $\bm{\hat{a}}_{lk}^{(j-1)}$.\\
\indent {\bf 4e)} If $\left(\sum\limits_{l=1}^L\sum\limits_{k=1}^K\left\Vert \bm{\bar{a}}_{lk}^{(j)}-\bm{\tilde{a}}_{lk}^{(j)}\right\Vert^2\right)\Big/ \left(\sum\limits_{l=1}^L\sum\limits_{k=1}^K\left\Vert \bm{\tilde{a}}_{lk}^{(j)}\right\Vert^2\right)\leq \epsilon^{\rm ADMM}$ where $\epsilon^{\rm ADMM}>0$ is a predefined threshold (which quantifies the convergence of the primal variables to each other), then set $\bm{a}_{lk}^{(i)}=\bm{\Omega}_{k}^{-1}\bm{\tilde{a}}_{lk}^{(j)}$ and continue with Step 5. Otherwise go to Step 4a.\\
{\bf 5)} If $\left|\sum\limits_{l=1}^L\sum\limits_{k=1}^K\left(\log_2\left(d_{lk}^{(i)}\right)-\log_2\left(d_{lk}^{(i-1)}\right)\right)\right|^2\Big/\left|\sum\limits_{l=1}^L\sum\limits_{k=1}^K\log_2\left(d_{lk}^{(i-1)}\right)\right|^2\leq \epsilon^{\rm WMMSE}$ where $\epsilon^{\rm WMMSE}>0$ is a predefined threshold (which quantifies the improvement in the sum SE), then stop with solution $\left\{\bm{a}_{lk}^{(i)}\right\}$. Otherwise continue with Step 2.
\vspace{-1.2cm}
\begin{center}
	\linethickness{0.15mm}
	\line(1,0){485}
\end{center}

Note that the sum SE is a performance metric that does not mind the individual performance of the users. To obtain the highest possible sum SE, the maximization problem in (\ref{eq:obj})-(\ref{eq:const1}) puts more emphasis on the users with good signal-to-noise ratio (SNR). Hence, it does not guarantee any user fairness. In those setups where the fairness is the main criteria, maximizing the minimum SE of the network can be selected as the optimization criteria instead \cite{interference_marzetta}. The major drawback of max-min fairness based optimization is that the users with the worst channel conditions will reduce the SE of other users severely. Another alternative scheme that provides a balance with these two extreme cases (sum and the worst SE-centric) is the proportional fairness, which we consider in the next section.

\section{LSFP Optimization for Proportional Fairness}
\label{sec:propFair}
In this section, we will design the LSFP weights $\{a_{rk}^l\}$ based on the proportional fairness criterion \cite{stationary,proportional1, proportional2,amin2}\footnote{In \cite{amin2}, a different name ``geometric-mean per-cell max-min fairness'' is used for conventional proportional fairness with a slightly modified objective.}, which aims to maximize the sum of logarithms of the individual SE of the users under per-BS transmit power constraints. This is an optimization objective that obtains a good balance between max-min fairness, which limits the network-wide performance by focusing on the users with the worst channel conditions, and sum SE maximization, which does not guarantee any user fairness. We have previously maximized the arithmetic mean of the user SEs and now we will maximize the geometric mean instead, where the corresponding optimization problem is cast as
\begin{align}
&\underset{\left\{\bm{a}_{lk}\right\}}{\text{maximize}} \ \ \ \sum_{l=1}^L\sum_{k=1}^K\ln\left(\log_2\left(1+\sinr_{lk}\right)\right) \label{eq:obj3} \\
&\text{subject to} \ \ \ \sum\limits_{k=1}^K\omega_{lk}\sum\limits_{r=1}^L|a_{rk}^l|^2\leq\rho_d, \ \ l=1,\ldots,L, \label{eq:const31} 
\end{align}
which is a non-convex problem where a global optimum solution with a polynomial time complexity is not guaranteed. In this paper, we will apply the weighted-MMSE formulation for sum SE maximization by using the results in \cite{stationary}. Please see Theorem 2 and the following discussion in \cite{stationary}. It says that we can obtain an equivalent weighted MMSE-type problem to (\ref{eq:obj3})-(\ref{eq:const31}) in the sense that they have the same global optimum solution:
\begin{align}
&\underset{\left\{\bm{a}_{lk},  \ u_{lk}, \ d_{lk}\geq 0 \right\}}{\text{minimize}} \ \ \ \sum_{l=1}^L\sum_{k=1}^K\Big(d_{lk}e_{lk}+f\left(g\left(d_{lk}\right)\right)-d_{lk}g\left(d_{lk}\right)\Big) \label{eq:obj5} \\
&\hspace{0.8cm}\text{subject to} \ \ \ \sum\limits_{k=1}^K\omega_{lk}\sum\limits_{r=1}^L|a_{rk}^l|^2\leq\rho_d, \ \ l=1,\ldots,L, \label{eq:const51} 
\end{align}
where $e_{lk}$ is as in \eqref{eq:elk}. The function $f(x)$ is defined as $f(x)=-\ln\left(-\ln\left(x\right)\right)$ for $0<x<1$. Its derivative function $f^{\prime}(x)$ is given by $f^{\prime}(x)=-1/\left(x\ln(x)\right)$, which is positive for $0<x<1$. The function $g(x)$ is the right inverse function of $f^{\prime}(x)$, i.e., $f^{\prime}\left(g(x)\right)=x$.

\begin{remark} By \cite[Theorem 2]{stationary} and the following discussion, the function $g(x)$ is assumed to be well-defined, and $f(x)$ should be strictly concave in the region of interest for this to happen. In \cite{stationary}, it is claimed that $f(x)=-\ln\left(-\ln\left(x\right)\right)$ strictly concave. However, a simple calculation shows that this is not the case for the entire region in $0<x<1$. The authors think that there is a redundant requirement regarding Theorem 2 in \cite{stationary}. As the following steps show, there is no need to evaluate the function $g(x)$ explicitly and we can simply assume $f^{\prime}\left(g(x)\right)=x$, which is valid only for one direction and can be obtained by some vector-valued function $g(x)$. Since it is not required to know $g(x)$ in the block coordinate descent algorithm steps, we will not go deep into the details, and continue with the scalar function $g(x)$ without loss of generality. 
\end{remark}
 The equivalency of two problems (\ref{eq:obj3})-(\ref{eq:const31}) and (\ref{eq:obj5})-(\ref{eq:const51}) follows from the fact that the optimum $d_{lk}$ for the above problem is obtained by equating the derivative of the objective function to zero (since the constraints are independent of $d_{lk}$):
\begin{align}
e_{lk}+f^{\prime}\left(g\left(d_{lk}\right)\right)g^{\prime}\left(d_{lk}\right)-g\left(d_{lk}\right)-d_{lk}g^{\prime}\left(d_{lk}\right)=0,
\end{align}
which is equivalent to
\begin{align}
e_{lk}=g(d_{lk}) \implies d_{lk}=f^{\prime}(e_{lk}),
\end{align}
where we have used $f^{\prime}\left(g\left(d_{lk}\right)\right)=d_{lk}$. When we insert the optimum $d_{lk}$ given above into the objective function in \eqref{eq:obj5}, we obtain 
\begin{align}
&\sum_{l=1}^L\sum_{k=1}^K\Big(f^{\prime}(e_{lk})e_{lk}+f\left(g\left(f^{\prime}(e_{lk})\right)\right)-f^{\prime}(e_{lk})g\left(f^{\prime}(e_{lk})\right)\Big)\nonumber \\
&=\sum_{l=1}^L\sum_{k=1}^K f(e_{lk})=\sum_{l=1}^L\sum_{k=1}^K -\ln\left(-\ln\left(e_{lk}\right)\right)=-\sum_{l=1}^L\sum_{k=1}^K\ln(\ln(1+\sinr_{lk})), \label{eq:obj6}
\end{align}
where we have used that $e_{lk}$ with optimum $u_{lk}$ is equal to $1/(1+\sinr_{lk})$. Minimizing the function in \eqref{eq:obj6} is equal to maximizing the proportional fairness metric in \eqref{eq:obj3}. Hence, we obtain the equivalent problem to (\ref{eq:obj3})-(\ref{eq:const31}) in (\ref{eq:obj5})-(\ref{eq:const51}). It can easily be shown that Theorem~\ref{the3} is also valid for the problem (\ref{eq:obj5})-(\ref{eq:const51}) and we can use the block coordinate descent Algorithm~1 to obtain a stationary point to this problem with closed-form updates. The only modification occurs in Step 3 of Algorithm~1 where $\left\{d_{lk}^{(i)}\right\}$ need to be updated as $d_{lk}^{(i)}=f^{\prime}(e_{lk}^{(i-1)})=-1/\left(e_{lk}^{(i-1)}\ln\left(e_{lk}^{(i-1)}\right)\right)$.\footnote{Note that $e_{lk}^{(i-1)}$ is equal to $1/(1+\sinr_{lk})<1$ by the previous optimum updates, and, hence, in terms of domain restrictions of the newly defined functions, there is no conflict.}  To avoid repetition, we are not presenting the steps of the block coordinate descent algorithm for proportional fairness here. In the simulations, we will compare the sum SE and proportional fairness maximization problems.
\section{Partial LSFP}

The aim of LSFP is to mitigate coherent interference resulting from pilot contamination as well as non-coherent interference with a proper power allocation. Implementation of LSFP requires BSs sharing not only the long-term channel statistics but also individual downlink data signals of the users, which may put heavy burden on the fronthaul links to the central  network controller as the number of cooperating BSs increases. One extreme case is single-layer precoding where the BSs only serve the users in their cells and the central network controller only needs user channel statistics to optimize the power allocation coefficients. In this scheme, the optimization algorithms in the previous sections are implemented by setting all the elements of $\bm{a}_{lk}$ except the $l^{\textrm{th}}$ one to zero. To reduce the fronthaul requirements for a large network, one option is to introduce \emph{partial LSFP} where some of the vectors $\bm{a}_{lk}$ have only one non-zero element, which is $a_{lk}^l$, which is determined by a predefined number. This can be motivated by the fact that LSFP is mainly useful for those users that are subject to high inter-cell interference. Hence, we don't need to pass around data for the cell-center users to other BSs. Note that the number of downlink symbols that are required to be sent to the central network controller to calculate the combined signals to be transmitted from each BS in \eqref{eq:vlk} is proportional to the number of LSFP vectors $\bm{a}_{lk}$ that have more than one non-zero element. For LSFP, $(\tau_c-\tau_p)LK$ downlink symbols in each coherence block are required to be shared with the central network controller and then the central network controller should calculate and send the combined signals in \eqref{eq:vlk} to each BS, which is again $(\tau_c-\tau_p)LK$ downlink symbols in each coherence block. On the other hand, for single-layer precoding no sharing of downlink data is needed. Let $1\leq N_D<LK$ denote the total number of the vectors $\bm{a}_{lk}$ that have more than one non-zero element for partial LSFP. Then the BSs are not required to share the downlink symbols of the users whose corresponding $\bm{a}_{lk}$ has only one non-zero element, i.e., $a_{lk}^l$, which represents the power allocation coefficient. The only thing  is that the central network controller sends the power allocation coefficient of user $k$ to BS $l$, which is only done when the long-term channel statistics change. Hence, the total number of downlink symbols to be shared in each coherence block is $(\tau_c-\tau_p)N_D$ for partial LSFP.

\begin{remark} The main burden on the fronthaul signaling is to send data, while sending around parameters like power allocation coefficients that only depend on the long-term statistics is almost negligible. Note that the power allocation coefficients to scale the local precoders at each BS are also shared in single-layer implementation. 
\end{remark}
   In the following, we propose two heuristics for selecting the vectors $\bm{a}_{lk}$ that have more than one non-zero element by using the long-term channel statistics information in \eqref{eq:blk} and \eqref{eq:Clk}. First, let $\mathcal{D}$ denote the set of BS and user index pairs $(l,k)$ corresponding to these $\{\bm{a}_{lk}\}$. For other indices that are not included in the set $\mathcal{D}$, all the elements except the $l$th one of the corresponding $\bm{a}_{lk}$ are set to zero. The number of elements in the set $\mathcal{D}$ is thus $\vert \mathcal{D}\vert=N_D$ where $N_D$ is a predefined value.

\subsection{Selection of Partial LSFP Indices Based on Only the Desired Signal Strength}
\label{sec:partial-lsfp-b}
Let us only focus on the desired signal strength that is represented by the vector $\bm{b}_{lk}$ for user $k$ in cell $l$ in \eqref{eq:SE1}. The power of $b_{lk}^l$, which is multiplied by the power allocation coefficient $a_{lk}^l$ in single-layer precoding, relative to the norm square of $\bm{b}_{lk}$ quantifies the level to what extent user $k$ in cell $l$ can benefit from LSFP. When it is small, we expect that invoking LSFP by including the weights other than $a_{lk}^l$, will improve the SE of that user. As a heuristic method, the BS and user index pairs $(l,k)$ in the set $\mathcal{D}$ can be determined by sorting the normalized power of the elements in $\bm{b}_{lk}$ and selecting the indices corresponding to the smallest values. For each BS and user pair $(l,k)$, we consider $|b_{lk}^l|^2/\Vert \bm{b}_{lk}\Vert^2$ and sort these values. Then, the set $\mathcal{D}$ is constructed by the BS and user index pairs  $(l,k)$ corresponding to the smallest $N_D>0$ values. We apply LSFP for the $N_D$ users with the smallest values since these are the UEs that are most affected by the surrounding BSs. 
\subsection{Selection of Partial LSFP Indices Based on the Desired Signal Strength and Interference}
\label{sec:partial-lsfp-c}
This heuristics method also takes the interference statistics that is represented by the elements of the matrices $\bm{C}_{lkk^{\prime}}$ for user $k$ in cell $l$ into account in addition to the vectors $\bm{b}_{lk}$, which represents the signal strengths. In an effort to maximize the SINR in \eqref{eq:SE1}, we can construct the following metric to be maximized
\begin{align}\label{eq:Slk}
&S_{lk}=\frac{\left|\bm{a}_{lk}^H\bm{b}_{lk}\right|^2}{\bm{a}_{lk}^H\bigg(\sum\limits_{\substack{r=1}}^L\sum\limits_{\substack{k^{\prime}=1}}^K\bm{C}_{rk^{\prime}k}\bigg)\bm{a}_{lk}}
\end{align}
where the numerator and denominator represent respectively the desired signal power of user $k$ in cell $l$ and the interference that it creates to other users in the network. $S_{lk}$ is maximized by the vector $\bm{a}_{lk}^{\star}=\bigg(\sum\limits_{\substack{r=1}}^L\sum\limits_{\substack{k^{\prime}=1}}^K\bm{C}_{rk^{\prime}k}\bigg)^{-1}\bm{b}_{lk}$ by Rayleigh quotient.\footnote{It can be shown that the matrix whose inverse is taken is non-singular using the definition of the matrices $\{\bm{C}_{rk^{\prime}k}\}$ in \eqref{eq:Clk}.} Using the same reasoning in the previous section, we can introduce a selection criteria by ordering the relative strength of $\left(a_{lk}^l\right)^{\star}$. Similar to the first method, the set $\mathcal{D}$ is constructed by the BS and user indices $(l,k)$ corresponding to the smallest $N_D>0$ values in $\left\{|\left(a_{lk}^l\right)^{\star}|^2/\Vert \bm{a}_{lk}^{\star}\Vert^2:l=1,\ldots,L,k=1,\ldots,K\right\}$. 

After selecting for which BS and user LSFP is implemented, we can solve the optimization problems by the proposed block descent algorithm with the non-zero elements of $\bm{a}_{lk}$, which are determined by the index pairs in $\mathcal{D}$. 
\section{Numerical Results}

In this section, we compare the downlink SE performance of several precoding and power allocation schemes with either LMMSE- or LS-based channel estimation. The schemes that are optimized using the sum SE maximization method in Section~\ref{sec:sumSE} are: 
\begin{itemize}
	\item {\bf LSFP-SumSE:} The proposed LSFP scheme (two-layer precoding).
			\item {\bf P-DS-LSFP-SumSE:} The proposed partial LSFP scheme in Section~\ref{sec:partial-lsfp-b} that is based on only the desired signal strength and $N_D=LK/2$ corresponding to the half fronthaul signaling load compared to LSFP.
		\item {\bf P-DS+Int-LSFP-SumSE:} The proposed partial LSFP scheme in Section~\ref{sec:partial-lsfp-c} that is based on both the desired signal strength and interference with $N_D=LK/2$.
		\item {\bf SLP-SumSE:} Standard single-layer precoding by setting the all the entries of the vector $\bm{a}_{lk}$ to zero except the $l$th entry. 
\end{itemize}
The LSFP and SLP schemes, which are optimized for proportional fairness maximization method in Section~\ref{sec:propFair} are called {\bf LSFP-PropFair} and {\bf SLP-PropFair}, respectively. For SLP schemes, each BS only transmit data to their own users. As a simple power allocation benchmark, we also consider the heuristic approach in \cite{giovanni}, where the downlink signal power of user $k$ in cell $l$ at its serving BS $l$ is proportional to $\sqrt{\mathbb{E}\left\{\Vert{\bf w}_{lk}\Vert^2\right\}}$ and total transmitted power from each BS is $\rho_d$. In the figures, this scheme is denoted by {\bf LPA} (local power allocation).

We consider mainly the Rician fading multi-cell setup in \cite{ozge_massive} which is based on the 3GPP model in \cite{channel}. Different from the setup in \cite{ozge_massive}, in our scenario, the phases of the LOS components are shifted randomly in every coherence block. There are $L=16$ cells in the network where each cell occupies a 250\,m$\times$250\,m square area with the BS at the center. The number of antennas at each BS is $M=200$. There are $K=8$ users in each cell. The uplink pilot power is $\eta=0.1$\,W and the maximum downlink transmit power is $\rho_d=10$\,W. The  bandwidth is 20\,MHz and the thermal noise variance is $\sigma^2=-96$\,dBm. The length of each coherence block is $\tau_c=200$ with $\tau_p=K=8$. We present the results of 100 different setups where the users are dropped in the cells uniformly with at least 20 m distance to the BSs in accordance with the urban microcell model in \cite{channel}. We consider a 11\,m height difference between the BSs and UEs in the path loss calculation. We assume the antennas of each BS are deployed in a uniform linear array (ULA) configuration with half-wavelength spacing and the deterministic part of the channel from user $k$ in cell $l$ to BS in the cell $r$ is
\begin{align}
{{\bf \bar{g}}}_{lk}^{r}=\sqrt{\beta_{lk}^{r \,\rm LOS}}\left[ \ 1 \ e^{j\pi\sin\left(\phi_{lk}^r\right)\cos\left(\psi_{lk}^r\right)} \ \ldots \  e^{j\pi\left(M-1\right)\sin\left(\phi_{lk}^r\right)\cos\left(\psi_{lk}^r\right)} \ \right]^T,
\end{align}
where $\beta_{lk}^{r \,\rm LOS}$ is the gain of the LOS part of the channel. The angles $\phi_{lk}^r$ and $\psi_{lk}^r$ are respectively the azimuth and elevation angles of user $k$ in cell $l$ with respect to BS $r$. The local scattering spatial correlation model in \cite[Section~2.6]{emil_book} is used for generating the correlation matrices $\{{\bf R}_{lk}^r\}$ with the approximate expression in \cite[Equation~(2.24)]{emil_book} and the effective azimuth angle $\arcsin\left(\sin\left(\phi_{lk}^r\right)\cos\left(\psi_{lk}^r\right)\right)$ is used to take the elevation angle into account.

The parameters for the solution accuracy in Algorithm~1 are selected as $\epsilon^{\rm ADMM}=\epsilon^{\rm WMMSE}=10^{-5}$ and the penalty parameter for the ADMM method is $\rho=0.2$ based on the empirical simulations.

\begin{figure}[t!]
	\hspace{-0.5cm}
	\begin{minipage}[t]{0.5\linewidth}
		\begin{center}
			\includegraphics[trim={1.9cm 0cm 2.9cm 1.1cm},clip,width=\textwidth]{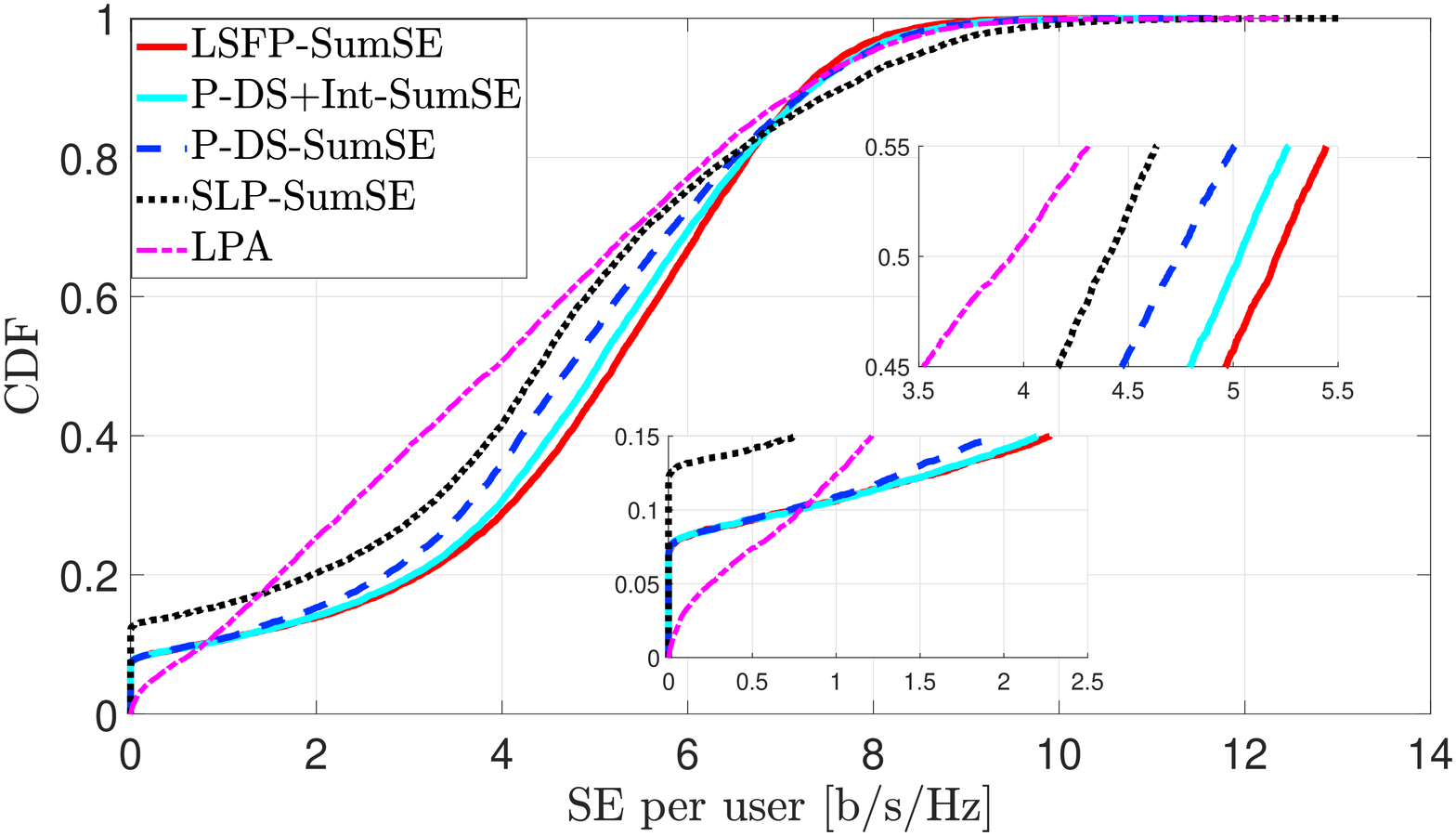}  \vspace{-8mm}
			\caption{SE per user for sum-SE maximization with the LS-based channel estimation.} \label{fig:1a}
		\end{center}
	\end{minipage}
	\hspace{0.5cm}
	\begin{minipage}[t]{0.5\linewidth}
		\begin{center}
			\includegraphics[trim={1.9cm 0cm 2.9cm 1.1cm},clip,width=\textwidth]{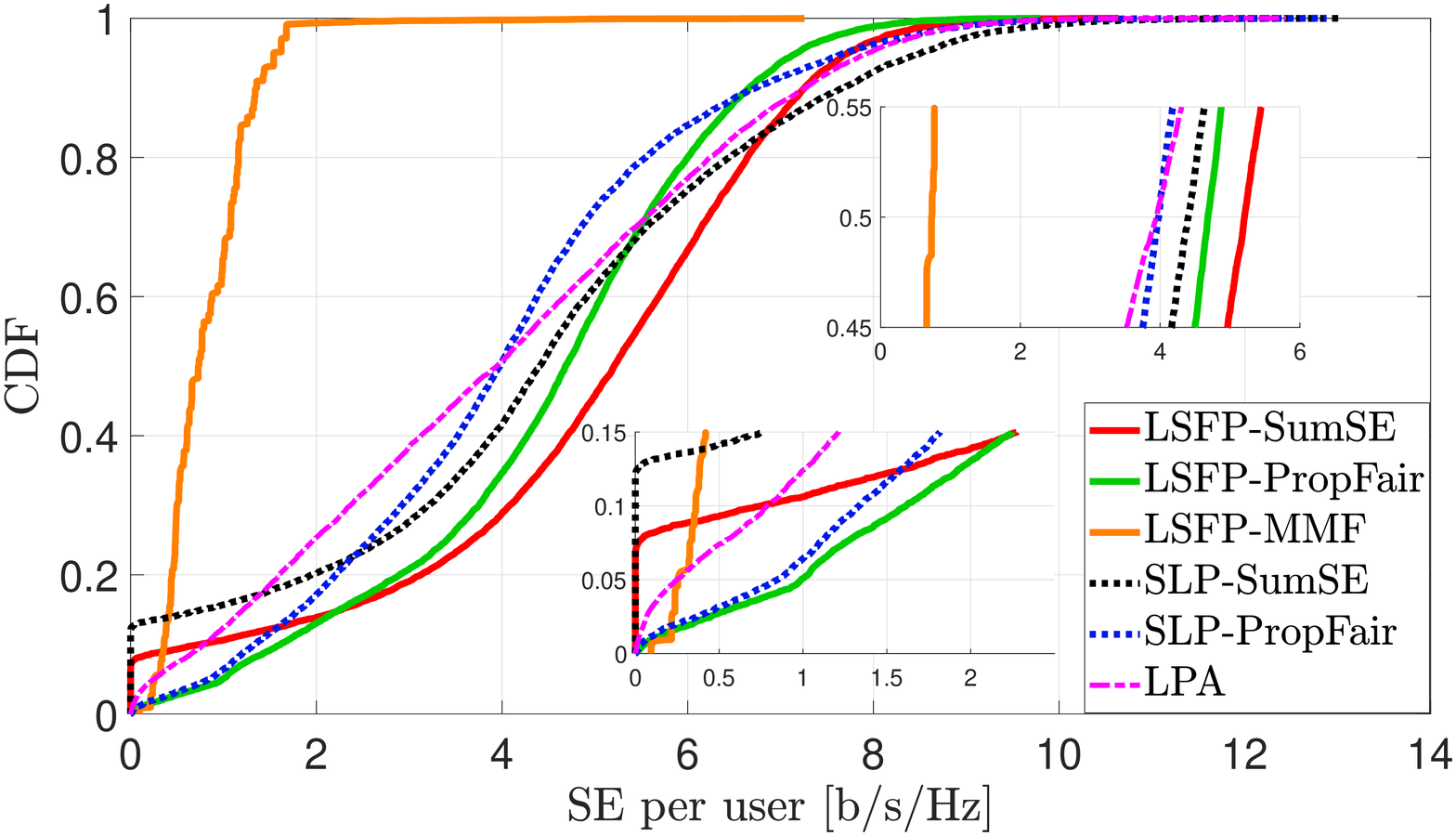} \vspace{-8mm}
			\caption{SE per user for different optimization criteria with the LS-based channel estimation.} \label{fig:1b}
		\end{center}
	\end{minipage} \vspace{-0.6cm}
\end{figure}

In Fig.~\ref{fig:1a}, we plot the cumulative distribution function (CDF) of the SE per user for the LSFP and SLP schemes that are optimized to maximize the sum SE, and the LPA. In this scenario, the local MR precoders are selected based on the LS-based channel estimation. In this figure and the following figures, we also present two zoomed versions of the main plot to quantify the gap between different schemes for near 90\% likely SE (where the CDF is 0.1) and the median SE (where the CDF is 0.5). The former one represents the minimum SE that the multi-cell network can provide to 90\% of the users, which represents user fairness since this value is determined by the users with relatively worse channel conditions. As Fig.~\ref{fig:1a} shows, LSFP improves the 90\% likely SE significantly in comparison to the SLP. However, the heuristic method LPA results in nearly the same SE at this point. As can be seen from the CDF where it is between 0.1 and 0.8, LSFP provides significant SE improvement compared to the other schemes. In fact, the median SE with LSFP-SumSE is 18\% and 32\% higher in comparison to SLP-SumSE and LPA, respectively. Moreover, both partial schemes P-DS+Int-LSFP and P-DS-LSFP perform very close to the LSFP in the lower part of the CDF curve with less fronthaul signaling load as quantified in Section~\ref{sec:partial-lsfp-b} and Section~\ref{sec:partial-lsfp-c}. However, at the median point, there is some performance loss in comparison to the full LSFP, which provides the highest median SE among all the schemes. Note that P-DS+Int-SumSE provides higher median SE than P-DS-SumSE from Fig.~\ref{fig:1a} by taking the interference statistics into account in selecting the indices for partial LSFP implementation.

To see the impact of fairness improvement by proportional fairness, we consider the same scenario as before in Fig.~\ref{fig:1b} by including LSFP-PropFair and SLP-PropFair. We also include the results of max-min fairness optimization with LSFP from \cite{interference_marzetta}, which is solved optimally by a bisection search over second-order cone programs. The results are denoted by LSFP-MMF (max-min fairness) in Fig.~\ref{fig:1b}. As it can be seen from the bottom zoomed figure, the PropFair schemes (both with LSFP and SLP) provide higher SE to the worst users and provide more fairness. However, the tradeoff occurs for higher SE values as can be seen from the median point where both schemes have less SE than their SumSE counterparts. Note that LSFP-MMF provides the highest SE for the worst users in the network as can be seen from the bottom part of the CDFs. However, this results in a huge performance loss for most of the users. In an effort to maximize the worst SE by not caring the others, LSFP-MMF does not even provide higher 95\% likely SE than the PropFair schemes. This shows that proportional fairness is more suitable than max-min fairness for both user fairness and reasonable performance for all the users. 

\begin{figure}[t!]
	\hspace{-0.5cm}
	\begin{minipage}[t]{0.5\linewidth}
		\begin{center}
			\includegraphics[trim={1.9cm 0cm 2.9cm 1.1cm},clip,width=\textwidth]{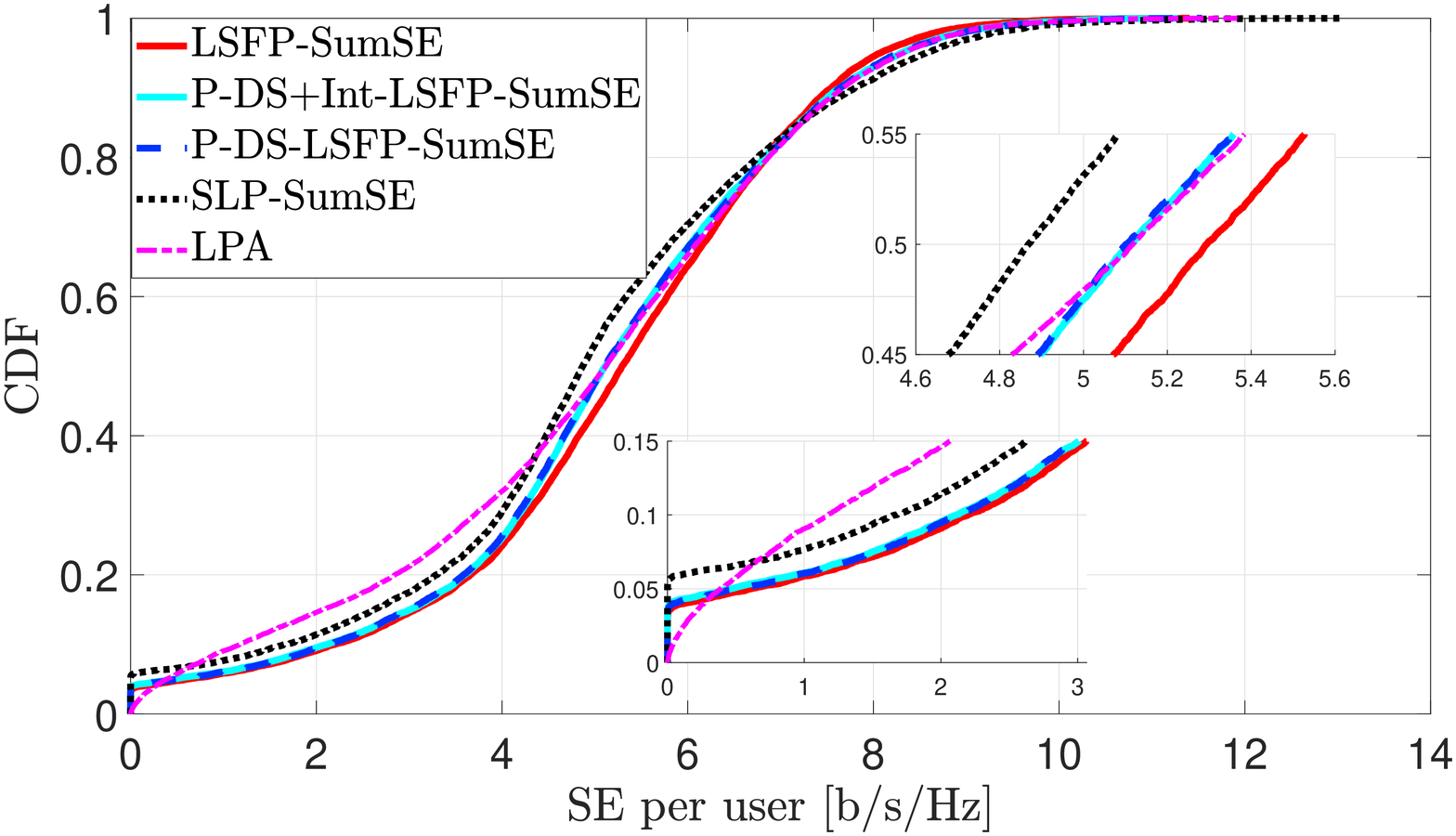}  \vspace{-8mm}
			\caption{SE per user for sum-SE maximization with the LMMSE-based channel estimation.} \label{fig:2a}
		\end{center}
	\end{minipage}
	\hspace{0.5cm}
	\begin{minipage}[t]{0.5\linewidth}
		\begin{center}
			\includegraphics[trim={1.9cm 0cm 2.9cm 1.1cm},clip,width=\textwidth]{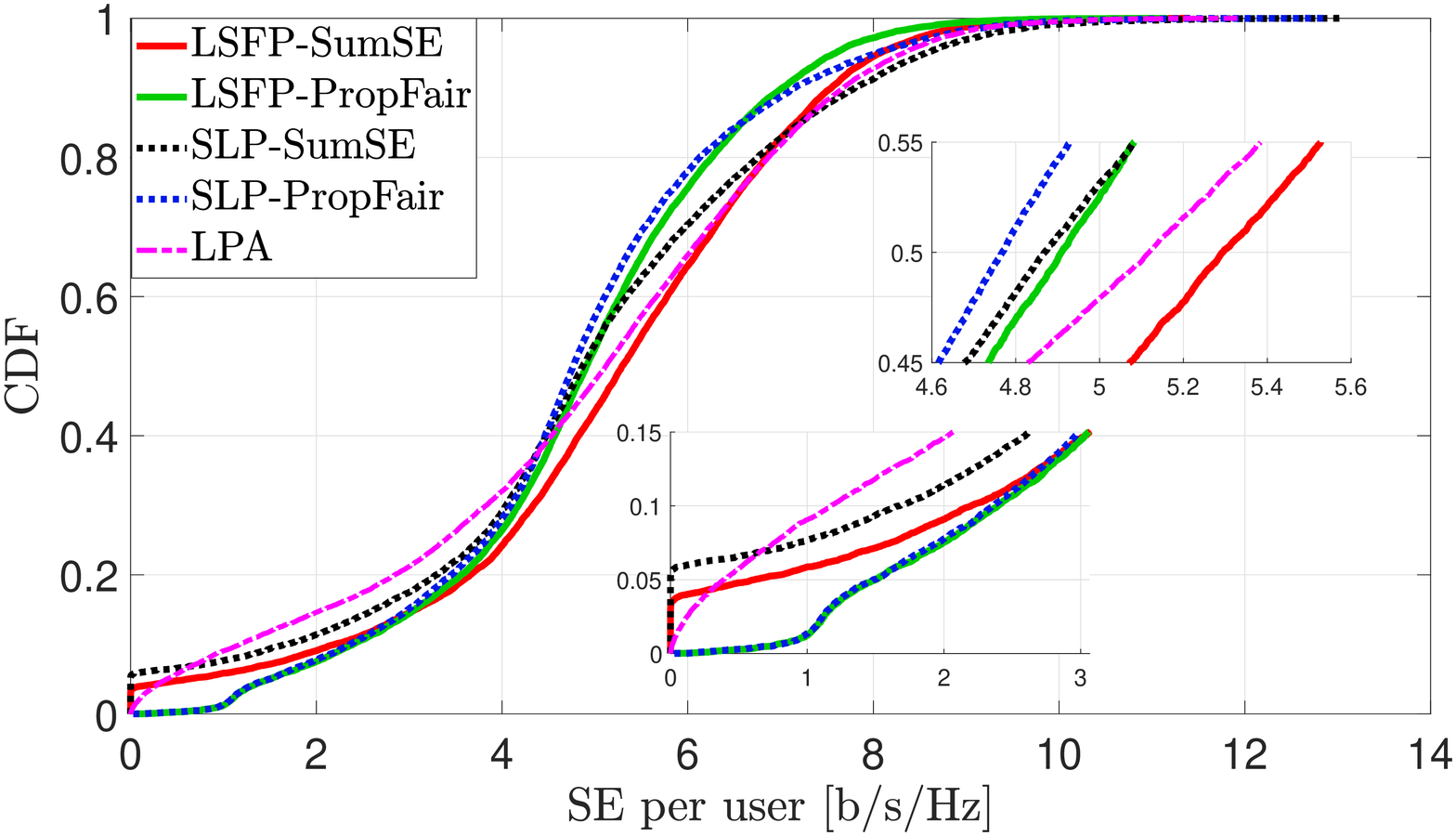} \vspace{-8mm}
			\caption{SE per user for different optimization criteria with the LMMSE-based channel estimation.} \label{fig:2b}
		\end{center}
	\end{minipage} \vspace{-0.6cm}
\end{figure}

In Fig.~\ref{fig:2a} and Fig.~\ref{fig:2b}, we repeat the previous experiment with LMMSE-based channel estimation. The main difference compared to the LS-based case, LSFP provides much higher 90\% likely SE than both SLP and LPA as can be seen from Fig.~\ref{fig:2a}. In fact, the 90\% likely SE with LSFP is 32\% and 87\% higher than SLP and LPA, respectively. Hence, it provides much more fairness. We see that both partial schemes P-DS+Int-LSFP and P-DS-LSFP perform very close to the LSFP. For LMMSE-based channel estimation, we see that the gap between these methods is negligible unlike the previous scenario with LS-based channel estimation. However, at the median point, their performances are close to the LPA. LSFP still provides the highest median SE among all the schemes.

As it can be seen from Fig.~\ref{fig:2b}, the PropFair schemes (both with LSFP and SLP) provide higher SE to the worst users and provide more fairness. However, the tradeoff occurs for higher SE values as can be seen from the median point where both schemes are much far from their SumSE counterparts. This median SE performance degradation is higher than before with LS-based channel estimation.

\begin{figure}[t!]
	\centering
	\includegraphics[trim={1.9cm 0cm 2.9cm 1.1cm},clip,width=8cm]{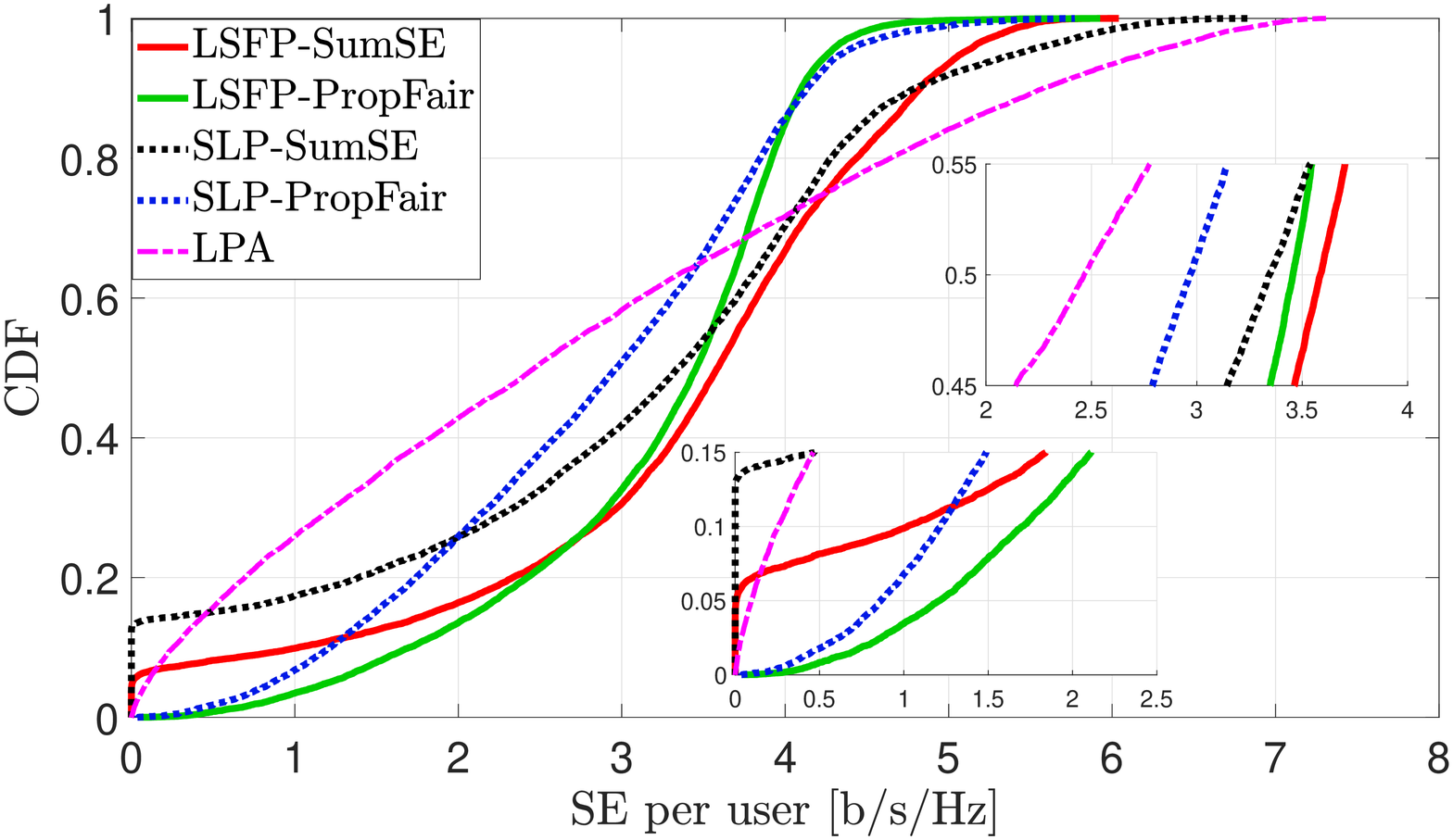}
	\vspace{-0.5cm}
	\caption{SE per user for spatially uncorrelated Rayleigh fading.}
	\label{fig:3x}
	\vspace{-0.3cm}
\end{figure}

The reason that the performance gap between LSFP and other standard single-layer precoding schemes is not as high as in LS-based channel estimation can be explained as follows. With LS-based channel estimation, the BSs are not able to resolve the channels between pilot sharing users since they do not utilize the spatial correlation between their antennas unlike LMMSE-based channel estimation. Hence, the improvement with LSFP becomes more significant since it has a larger room to suppress the inter-cell interference. If the channels follow spatially uncorrelated Rayleigh fading, then LMMSE- and LS-based channel estimates are the scaled version of each other. In this case, there do not exist any correlation among the BS antennas and any LOS paths. Hence, the BSs are not as successful as in spatially correlated fading in resolving different user channels and we expect a higher performance improvement this scenario. To see this effect, we repeat the previous experiment with spatially uncorrelated Rayleigh fading and plot the results in Fig.~\ref{fig:3x}. Now, the performance improvement with LSFP is higher compared to all the results before with spatially correlated Rician fading. In particular, LSFP-SumSE provides approximately 1\,b/s/Hz 90\% likely SE that is four times achieved with LPA. On the other hand, SLP-SumSE results in almost zero SE at this point. LSFP-PropFair provides much more 90\% likely SE, i.e., around 1.7\,b/s/Hz, which is 40\% higher than SLP-PropFair.

\begin{figure}[t!]
	\hspace{-0.5cm}
	\begin{minipage}[t]{0.5\linewidth}
		\begin{center}
			\includegraphics[trim={1.9cm 0cm 2.9cm 1.1cm},clip,width=\textwidth]{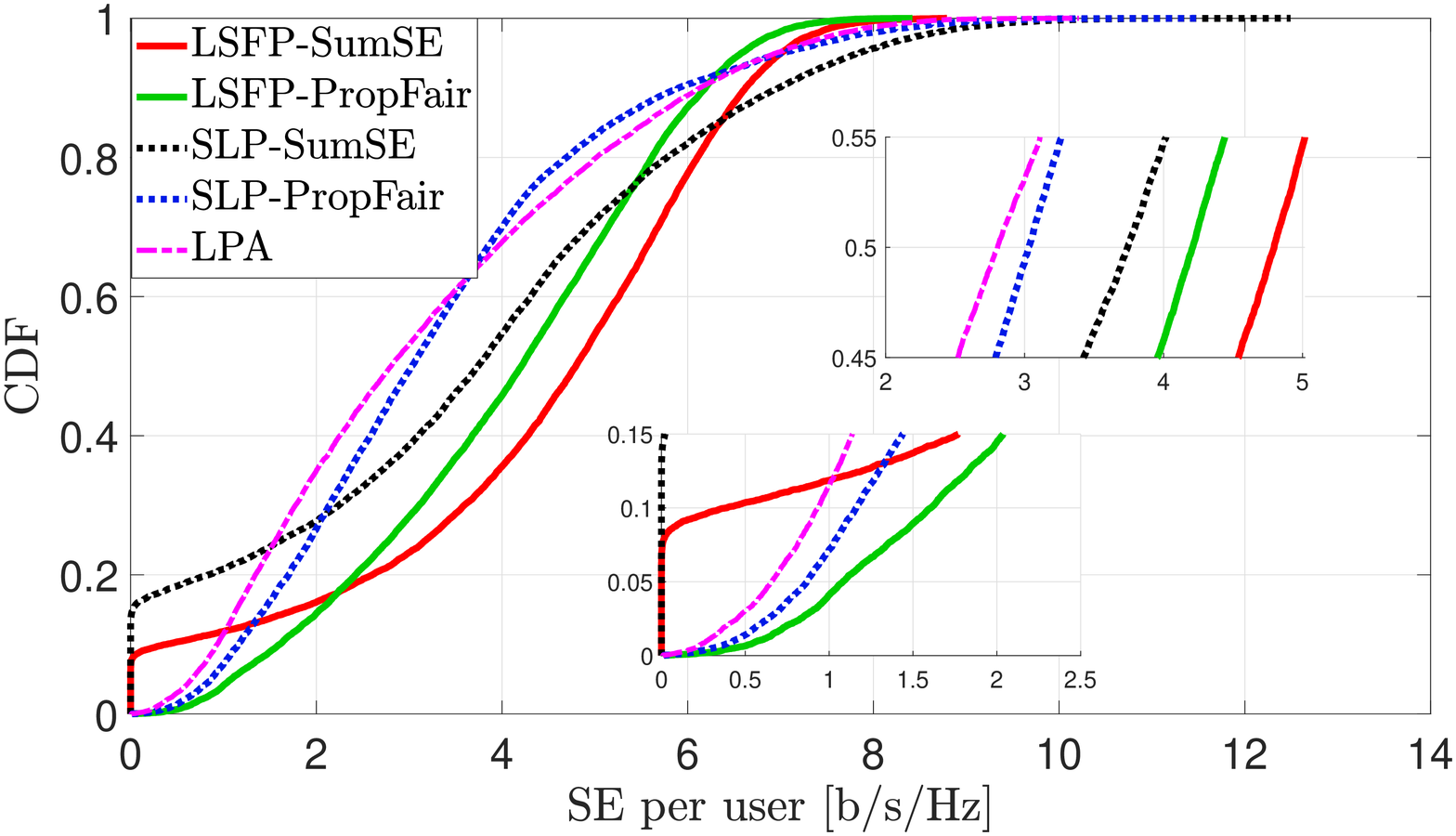}  \vspace{-8mm}
			\caption{SE per user for with the smaller cell size and LS-based channel estimation.} \label{fig:4a}
		\end{center}
	\end{minipage}
	\hspace{0.5cm}
	\begin{minipage}[t]{0.5\linewidth}
		\begin{center}
			\includegraphics[trim={1.9cm 0cm 2.9cm 1.1cm},clip,width=\textwidth]{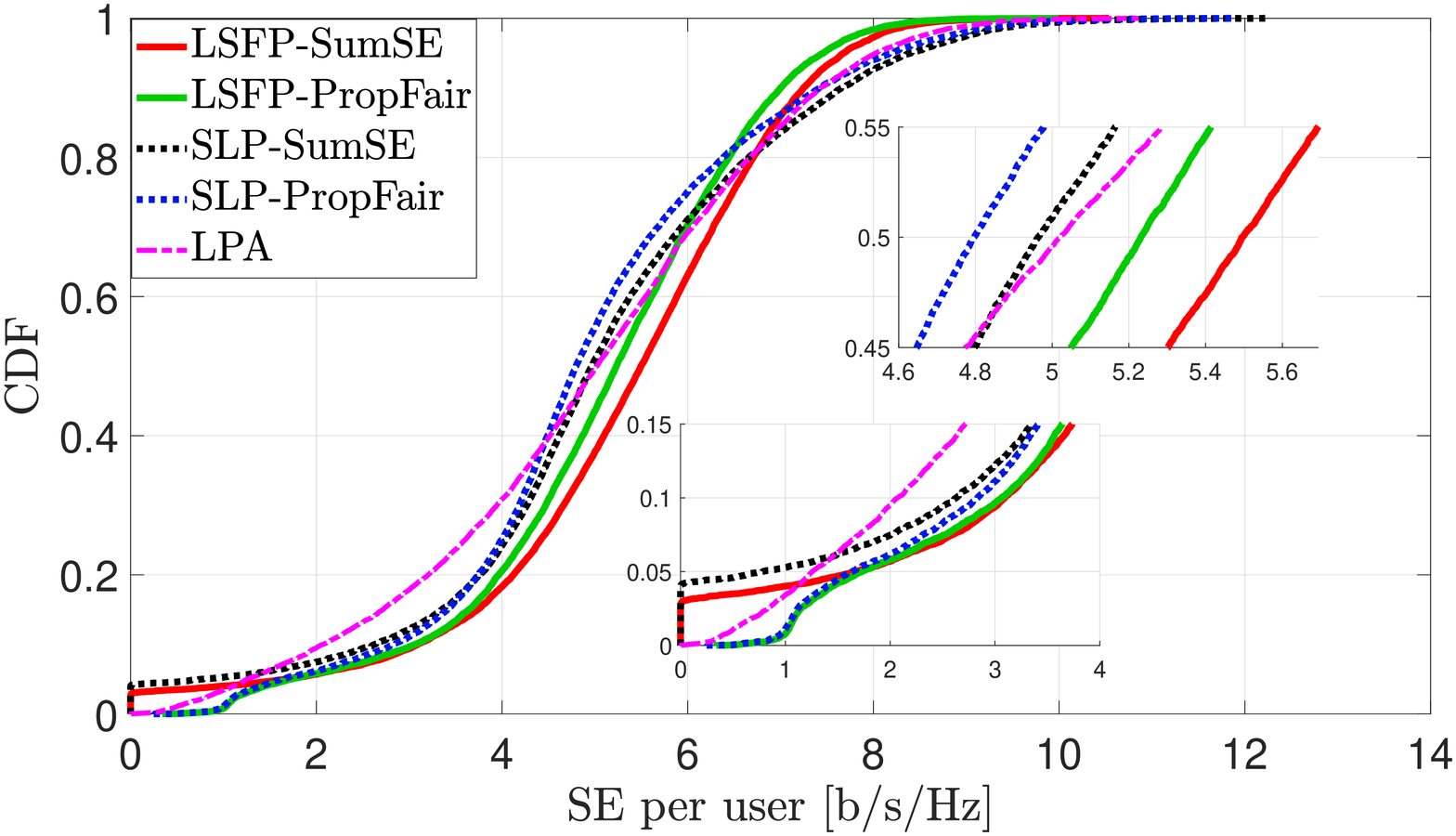} \vspace{-8mm}
			\caption{SE per user for with the smaller cell size and LMMSE-based channel estimation.} \label{fig:4b}
		\end{center}
	\end{minipage} \vspace{-0.6cm}
\end{figure}

As a final scenario, we consider a smaller cell size setup where each of $L=16$ cells occupies a 150\,m$\times$150\,m square area. All other parameters are the same with spatially correlated Rician fading model except for the uplink pilot power and the maximum downlink transmit power, which are lowered to $\eta=0.05$\,W and $\rho_d=5$\,W due to the reduced cell size. Fig.~\ref{fig:4a} and Fig.~\ref{fig:4b} present the CDF of the SE per user with LS- and LMMSE-based channel estimation, respectively. From both figures, it is obvious that the SE gap between LSFP and single-layer precoding schemes SLP and LPA is higher in comparison to the previous larger cell size that is 250\,m$\times$250\,m. From Fig.~\ref{fig:4a} where LS-based channel estimation is used for the MR local preocoding, the improvement of LSFP over other methods is much higher. In fact, the median SE with LSFP-SumSE is 28\% and 71\% higher than SLP-SumSE and LPA, respectively. LSFP-SumSE also provides higher median SE than LSFP-PropFair with less gain. However, 90\% likely SE with LSFP-PropFair is more than 3 times than that of LSFP-SumSE. From Fig.~\ref{fig:4b}, we also see that both the median and 90\% likely SE are higher with LSFP schemes. Again we note that the PropFair schemes provide the highest fairness for the worst users with an inevitable performance loss at the above parts of the CDF, i.e., at the median point.

\section{Conclusion}

In this paper, we have considered LSFP and compared its performance with several benchmarks in a realistic Rician fading environment where the LOS components of the channels are corrupted  by random phase shifts. We have proposed two efficient algorithms to optimize the LSFP weights at the central network controller to maximize sum SE and proportional fairness. The first observation is that LSFP is useful especially in the environments where pilot contamination is severe. It improves the SE of the worst users more than the others in the network.  Using a lower quality channel estimator, such as LS, results in a case where LSFP provides significantly higher SE than the single-layer precoding compared to the case of better channel estimators, such as LMMSE, which are able to resolve the channels under pilot contamination to certain extent. For a relatively smaller cell size, the improvement with LSFP is more substantial where a pilot contamination plays a major role. 

Sum SE and proportional fairness are shown to be two competing performance metrics for both two-layer and single-layer precoding schemes. Although proportional fairness maximization results in a great improvement in the SE of the worst users in the network, most of other users benefit more with sum SE maximization. Regarding 95\% and 90\% likely SE, proportional fairness maximization is even better than max-min fairness. In fact, max-min fairness only maximizes the SE of a few users among the worst ones, but has a substantial performance drop for most of the users compared to the other performance metrics.  

In addition, two simple partial LSFP schemes are proposed to determine for which BS and user pairs LSFP will be beneficial in an aim to take advantage of two-layer precoding with less fronthaul signaling requirements. For the users with worse channel gains, the partial LSFP schemes have very close performance to the full LSFP. For the other users, although there is a performance gap in general, partial LSFP schemes are better than single-layer implementation. Determining for which BS and user pairs LSFP is likely to be more useful is of great importance. For LS-based channel estimator, taking the long-term interference characteristics into account in the selection criterion results in better performance compared to the case with just considering desired signal strength. 

\appendices
\section{Useful Lemmas}
\begin{lemma}\label{applemma1}
	\cite[Lemma 2]{emil_nonideal}. Consider the  random vector ${\bf u}\in \mathbb{C}^M$ that is distributed as ${\bf u} \sim \mathcal{N}_{\mathbb{C}}\left({\bf 0}_M,{\bf A}\right)$. For a deterministic matrix ${\bf B}\in \mathbb{C}^{M\times M}$, it holds that
	\begin{align}
	&\mathbb{E}\left\{\left|{\bf u}^H{\bf B}{\bf u}\right|^2\right\}= \left|\tr\left({\bf A}{\bf B}\right)\right|^2+\tr\left({\bf A}{\bf B}{\bf A}{\bf B}^H\right).
	\end{align}
\end{lemma}

\begin{lemma}\label{applemma2}
	Consider the vectors ${\bf x}=e^{j\theta_x}{\bf \bar{x}}+{\bf \widetilde{x}} \in \mathbb{C}^{M}$ and ${\bf y}={\bf B}{\bf x}+{\bf z} \in \mathbb{C}^{M}$, where ${\bf \bar{x}} \in \mathbb{C}^{M}$ and ${\bf B}\in \mathbb{C}^{M\times M}$ are deterministic and $\theta_x$ is uniformly distributed on the interval $[0,2\pi)$. ${\bf \widetilde{x}}$ is independent of $\theta_x$ with ${\bf \widetilde{x}} \sim \mathcal{N}_{\mathbb{C}}({\bf 0}_M,{\bf A})$. ${\bf z} \in \mathbb{C}^{M}$ is a random vector independent of ${\bf x}$ and has zero-mean. Let ${\bf C}_y$ denote the covariance matrix of ${\bf y}$. Then, the following holds:
	\begin{align}
	&\mathbb{E}\left\{{\bf y}^H{\bf x}\right\}=\tr\left({\bf B}^H({\bf A}+{\bf \bar{x}}{\bf \bar{x}}^H)\right) \label{eq:applemma2-1a} \\
	&\mathbb{E}\left\{\left|{\bf y}^H{\bf x}\right|^2\right\}=\left|\tr\left({\bf A}{\bf B}\right)\right|^2+2\Re\left\{{\bf \bar{x}}^H{\bf B}{\bf \bar{x}}\tr\left({\bf B}^H{\bf A}\right)\right\}+\tr\left({\bf C}_y\left({\bf A}+{\bf \bar{x}}{\bf \bar{x}}^H\right)\right). \label{eq:applemma2-2a}
	\end{align}
	\begin{IEEEproof} Compute $\mathbb{E}\left\{{\bf y}^H{\bf x}\right\}$ as
		\begin{align}
		\mathbb{E}\left\{{\bf y}^H{\bf x}\right\}=\mathbb{E}\left\{{\bf x}^H{\bf B}^H{\bf x}\right\}+\mathbb{E}\left\{{\bf z}^H{\bf x}\right\}\stackrel{(a)}=\tr\left({\bf B}^H\mathbb{E}\left\{{\bf x}{\bf x}^H\right\}\right)\stackrel{(b)}=\tr\left({\bf B}^H\left({\bf A}+{\bf \bar{x}}{\bf \bar{x}}^H\right)\right),
		\end{align}
		where we used the independence of zero-mean ${\bf z}$ and ${\bf x}$ in $(a)$ and $\mathbb{E}\left\{{\bf x}{\bf x}^H\right\}={\bf A}+{\bf \bar{x}}{\bf \bar{x}}^H$ in $(b)$.
		
		Let us compute now $\mathbb{E}\left\{\left|{\bf y}^H{\bf x}\right|^2\right\}$ as
		\begin{align}
		\mathbb{E}\left\{\left|{\bf y}^H{\bf x}\right|^2\right\}=&\mathbb{E}\left\{\left({\bf x}^H{\bf B}^H+{\bf z}^H\right){\bf x}{\bf x}^H\left({\bf B}{\bf x}+{\bf z}\right)\right\}\stackrel{(a)}=\mathbb{E}\left\{{\bf x}^H{\bf B}^H{\bf x}{\bf x}^H{\bf B}{\bf x}\right\}+\mathbb{E}\left\{{\bf z}^H{\bf x}{\bf x}^H{\bf z}\right\} \nonumber\\
		\stackrel{(b)}=&{\bf \bar{x}}^H{\bf B}^H{\bf \bar{x}}{\bf \bar{x}}^H{\bf B}{\bf \bar{x}}+\mathbb{E}\left\{{\bf \bar{x}}^H{\bf B}^H{\bf \bar{x}}{\bf \widetilde{x}}^H{\bf B}{\bf \widetilde{x}}\right\}+\mathbb{E}\left\{{\bf \bar{x}}^H{\bf B}^H{\bf \widetilde{x}}{\bf \widetilde{x}}^H{\bf B}{\bf \bar{x}}\right\}\nonumber\\
		&+\mathbb{E}\left\{{\bf \widetilde{x}}^H{\bf B}^H{\bf \bar{x}}{\bf \bar{x}}^H{\bf B}{\bf \widetilde{x}}\right\}+\mathbb{E}\left\{{\bf \widetilde{x}}^H{\bf B}^H{\bf \widetilde{x}}{\bf \bar{x}}^H{\bf B}{\bf \bar{x}}\right\}+\mathbb{E}\left\{{\bf \widetilde{x}}^H{\bf B}^H{\bf \widetilde{x}}{\bf \widetilde{x}}^H{\bf B}{\bf \widetilde{x}}\right\}\nonumber \\
		&+\tr\left(\mathbb{E}\left\{{\bf z}{\bf z}^H\right\}\left({\bf A}+{\bf \bar{x}}{\bf \bar{x}}^H\right)\right) \nonumber \\
		\stackrel{(c)}=&{\bf \bar{x}}^H{\bf B}^H{\bf \bar{x}}{\bf \bar{x}}^H{\bf B}{\bf \bar{x}}+{\bf \bar{x}}^H{\bf B}^H{\bf \bar{x}}\tr\left({\bf A}{\bf B}\right)+{\bf \bar{x}}^H{\bf B}^H{\bf A}{\bf B}{\bf \bar{x}}+{\bf \bar{x}}^H{\bf B}{\bf A}{\bf B}^H{\bf \bar{x}}\nonumber\\
		&+{\bf \bar{x}}^H{\bf B}{\bf \bar{x}}\tr\left({\bf B}^H{\bf A}\right)+\left|\tr\left({\bf A}{\bf B}\right)\right|^2+\tr\left({\bf A}{\bf B}{\bf A}{\bf B}^H\right)\nonumber \\
		&+\tr\left(\left({\bf C}_y-{\bf B}\left({\bf A}+{\bf \bar{x}}{\bf \bar{x}}^H\right){\bf B}^H\right)\left({\bf A}+{\bf \bar{x}}{\bf \bar{x}}^H\right)\right),\label{eq:applemma2-2b}
		\end{align}
		where we used the independence of zero-mean ${\bf z}$ and ${\bf x}$ in $(a)$ and $(b)$. We have written all the non-zero individual terms of $\mathbb{E}\left\{{\bf x}^H{\bf B}^H{\bf x}{\bf x}^H{\bf B}{\bf x}\right\}$ separately by noting that $\theta_x$ is independent of ${\bf \widetilde{x}}$ and circular symmetry of ${\bf \tilde{x}}$ in $(b)$. We have used the cyclic shift property of trace and Lemma~\ref{applemma1}  together with $\mathbb{E}\left\{{\bf z}{\bf z}^H\right\}={\bf C}_y-{\bf B}\left({\bf A}+{\bf \bar{x}}{\bf \bar{x}}^H\right){\bf B}^H$ in $(c)$. After arranging the terms in \eqref{eq:applemma2-2b}, we obtain the result in \eqref{eq:applemma2-2a}.
	\end{IEEEproof}
\end{lemma}

\section{Proof of Theorem~\ref{the1}\label{the1_proof}}
Let us compute the expectations in Theorem 1 one by one.

1) Compute $b_{lk}^{r}=\mathbb{E}\left\{\left({\bf \hat{g}}_{rk}^r\right)^H{\bf g}_{lk}^{r}\right\}$ by using Lemma~\ref{applemma2} with ${\bf y}={\bf \hat{g}}_{rk}^r$, ${\bf x}={\bf g}_{lk}^{r}$, ${\bf \bar{x}}={\bf \bar{g}}_{lk}^{r}$, ${\bf A}={\bf R}_{lk}^r$, ${\bf B}=\tau_p\eta{\bf \overline{R}}_{rk}^r{\bf \Psi}_{rk}^{-1}$ as
\begin{align}
&b_{lk}^{r}=\tau_p\eta\tr\left({\bf \Psi}_{rk}^{-1}{\bf \overline{R}}_{rk}^{r}{\bf \overline{R}}_{lk}^{r}\right). \label{eq:blkr1app} 
\end{align}
2) Compute $c_{lkk}^{rr}=\mathbb{E}\left\{\left({\bf \hat{g}}_{rk}^r\right)^H{\bf g}_{lk}^{r}\left({\bf g}_{lk}^{r}\right)^H{\bf \hat{g}}_{rk}^r\right\}$ by using Lemma~\ref{applemma2} with ${\bf y}={\bf \hat{g}}_{rk}^r$, ${\bf x}={\bf g}_{lk}^{r}$, ${\bf \bar{x}}={\bf \bar{g}}_{lk}^{r}$, ${\bf A}={\bf R}_{lk}^r$, ${\bf B}=\tau_p\eta{\bf \overline{R}}_{rk}^r{\bf \Psi}_{rk}^{-1}$, and  ${\bf C}_y=\tau_p\eta{\bf \overline{R}}_{rk}^r{\bf \Psi}_{rk}^{-1}{\bf \overline{R}}_{rk}^r$ as
\begin{align}
&c_{lkk}^{rr}=\tau_p^2\eta^2\left|\tr\left({\bf R}_{lk}^{r}{\bf \overline{R}}_{rk}^{r}{\bf \Psi}_{rk}^{-1}\right)\right|^2+2\tau_p^2\eta^2\Re\left\{\left({\bf \bar{g}}_{lk}^r\right)^H{\bf \overline{R}}_{rk}^r{\bf \Psi}_{rk}^{-1}{\bf \bar{g}}_{lk}^r\tr\left({\bf \Psi}_{rk}^{-1}{\bf \overline{R}}_{rk}^r{\bf R}_{lk}^r\right)\right\}\nonumber\\
&\hspace{1cm}+\tau_p\eta\tr\left({\bf \overline{R}}_{rk}^r{\bf \Psi}_{rk}^{-1}{\bf \overline{R}}_{rk}^r{\bf \overline{R}}_{lk}^r\right) \label{eq:clkk-rr1app}.
\end{align}
3) Compute $c_{lkk}^{rn}=\mathbb{E}\left\{\left({\bf \hat{g}}_{rk}^r\right)^H{\bf g}_{lk}^{r}\left({\bf g}_{lk}^{n}\right)^H{\bf \hat{g}}_{nk}^n\right\}$ for $r\neq n$ as
\begin{align}
&c_{lkk}^{rn}=\mathbb{E}\left\{\left({\bf \hat{g}}_{rk}^r\right)^H{\bf g}_{lk}^{r}\right\}\mathbb{E}\left\{\left({\bf g}_{lk}^{n}\right)^H{\bf \hat{g}}_{nk}^n\right\}=b_{lk}^r\left(b_{lk}^n\right)^{*}, \ \ \ r\neq n, \label{eq:clkk-rn1app} 
\end{align}
where we have used the independence of channels and channel estimates corresponding to  BS $r$ with those corresponding to BS $n\neq r$ and the definition of $b_{lk}^r$ in the first step of the proof.

4) Compute $c_{lkk^{\prime}}^{rr}=\mathbb{E}\left\{\left({\bf \hat{g}}_{rk^{\prime}}^r\right)^H{\bf g}_{lk}^{r}\left({\bf g}_{lk}^{r}\right)^H{\bf \hat{g}}_{rk^{\prime}}^r\right\}$ for $k^{\prime} \neq k$ as
\begin{align}
&c_{lkk^{\prime}}^{rr}=\tr\left(\mathbb{E}\left\{{\bf \hat{g}}_{rk^{\prime}}^r\left({\bf \hat{g}}_{rk^{\prime}}^r\right)^H\right\}\mathbb{E}\left\{{\bf g}_{lk}^{r}\left({\bf g}_{lk}^{r}\right)^H\right\}\right)=\tau_p\eta\tr\left({\bf \overline{R}}_{rk^{\prime}}^r{\bf \Psi}_{rk^{\prime}}^{-1}{\bf \overline{R}}_{rk^{\prime}}^r{\bf \overline{R}}_{lk}^r\right), \ \ \  k^{\prime} \neq k, \label{eq:clkk2-rr1app} 
\end{align}
where we used the independence of channels and channel estimates for users that have different pilot sequences and \eqref{eq:covhath}.

5) Compute $c_{lkk^{\prime}}^{rn}=\mathbb{E}\left\{\left({\bf \hat{g}}_{rk^{\prime}}^r\right)^H{\bf g}_{lk}^{r}\left({\bf g}_{lk}^{n}\right)^H{\bf \hat{g}}_{nk^{\prime}}^n\right\}$ for $k^{\prime} \neq k$ and  $r \neq n$ as
\begin{align}
&c_{lkk^{\prime}}^{rn}=\mathbb{E}\left\{\left({\bf \hat{g}}_{rk^{\prime}}^r\right)^H\right\}\mathbb{E}\left\{{\bf g}_{lk}^{r}\right\}\mathbb{E}\left\{\left({\bf g}_{lk}^{n}\right)^H\right\}\mathbb{E}\left\{{\bf \hat{g}}_{nk^{\prime}}^n\right\}=0, \ \ \ k^{\prime} \neq k, \ r \neq n, \label{eq:clkk2-rn1app}
\end{align}
where we used the independence of zero-mean channels and channel estimates corresponding to different BSs and users with different pilot sequences.
\section{Proof of Theorem~\ref{the2}\label{the2_proof}}
Let us compute the expectations in Theorem 2 in the sequel. \\
1) Compute $b_{lk}^{r}=\mathbb{E}\left\{{\bf z}_{rk}^H{\bf g}_{lk}^{r}\right\}$ by using Lemma~\ref{applemma2} with ${\bf y}={\bf z}_{rk}$, ${\bf x}={\bf g}_{lk}^{r}$, ${\bf \bar{x}}={\bf \bar{g}}_{lk}^{r}$, ${\bf A}={\bf R}_{lk}^r$, ${\bf B}=\sqrt{\tau_p\eta}{\bf I}_M$ as
\begin{align}
b_{lk}^{r}=&\sqrt{\tau_p\eta}\tr\left({\bf \bar{g}}_{lk}^r\left({\bf \bar{g}}_{lk}^r\right)^H+{\bf R}_{lk}^{r}\right)\label{eq:blkr2app}.
\end{align}
2) Compute $c_{lkk}^{rr}=\mathbb{E}\left\{{\bf z}_{rk}^H{\bf g}_{lk}^{r}\left({\bf g}_{lk}^{r}\right)^H{\bf z}_{rk}\right\}$ by using Lemma~\ref{applemma2} with ${\bf y}={\bf z}_{rk}$, ${\bf x}={\bf g}_{lk}^{r}$, ${\bf \bar{x}}={\bf \bar{g}}_{lk}^{r}$, ${\bf A}={\bf R}_{lk}^r$, ${\bf B}=\sqrt{\tau_p\eta}{\bf I}_M$, and  ${\bf C}_y={\bf \Psi}_{rk}$  as
\begin{align}
c_{lkk}^{rr}&=\tau_p\eta\left(\tr\left({\bf R}_{lk}^{r}\right)\right)^2+2\tau_p\eta\left({\bf \bar{g}}_{lk}^r\right)^H{\bf \bar{g}}_{lk}^r\tr\left({\bf R}_{lk}^{r}\right)+\tr\left({\bf \Psi}_{rk}\left({\bf \bar{g}}_{lk}^r\left({\bf \bar{g}}_{lk}^r\right)^H+{\bf R}_{lk}^{r}\right)\right) \label{eq:clkk-rr2app}, 
\end{align}
3) Compute $c_{lkk}^{rn}=\mathbb{E}\left\{{\bf z}_{rk}^H{\bf g}_{lk}^{r}\left({\bf g}_{lk}^{n}\right)^H{\bf z}_{nk}\right\}$ for $r\neq n$ as
\begin{align}
&c_{lkk}^{rn}=\mathbb{E}\left\{{\bf z}_{rk}^H{\bf g}_{lk}^{r}\right\}\mathbb{E}\left\{\left({\bf g}_{lk}^{n}\right)^H{\bf z}_{nk}\right\}=b_{lk}^r\left(b_{lk}^n\right)^{*}, \ \ \ r\neq n, \label{eq:clkk-rn2app} 
\end{align}
where we have used the independence of channels and sufficient statistics corresponding to  BS $r$ with those corresponding to BS $n \neq r$ and the definition of $b_{lk}^r$ in the first step of the proof.

4) Compute $c_{lkk^{\prime}}^{rr}=\mathbb{E}\left\{{\bf z}_{rk^{\prime}}^H{\bf g}_{lk}^{r}\left({\bf g}_{lk}^{r}\right)^H{\bf z}_{rk^{\prime}}\right\}$ for $k^{\prime} \neq k$ as
\begin{align}
&c_{lkk^{\prime}}^{rr}=\tr\left(\mathbb{E}\left\{{\bf z}_{rk^{\prime}}{\bf z}_{rk^{\prime}}^H\right\}\mathbb{E}\left\{{\bf g}_{lk}^{r}\left({\bf g}_{lk}^{r}\right)^H\right\}\right)=\tr\left({\bf \Psi}_{rk^{\prime}}{\bf \overline{R}}_{lk}^{r}\right), \ \ \  k^{\prime} \neq k, \label{eq:clkk2-rr2app} 
\end{align}
where we used the independence of channels and sufficient statistics for users that have different pilot sequences and \eqref{eq:Psi}.

5) Compute $c_{lkk^{\prime}}^{rn}=\mathbb{E}\left\{{\bf z}_{rk^{\prime}}^H{\bf g}_{lk}^{r}\left({\bf g}_{lk}^{n}\right)^H{\bf z}_{nk^{\prime}}\right\}$ for $k^{\prime} \neq k$ and  $r \neq n$ as
\begin{align}
&c_{lkk^{\prime}}^{rn}=\mathbb{E}\left\{{\bf z}_{rk^{\prime}}^H\right\}\mathbb{E}\left\{{\bf g}_{lk}^{r}\right\}\mathbb{E}\left\{\left({\bf g}_{lk}^{n}\right)^H\right\}\mathbb{E}\left\{{\bf z}_{nk^{\prime}}\right\}=0, \ \ \ k^{\prime} \neq k, \ r \neq n, \label{eq:clkk2-rn2app}
\end{align}
where we used the independence of zero-mean channels and sufficient statistics corresponding to different BSs and users with different pilot sequences.

\ifCLASSOPTIONcaptionsoff
  \newpage
\fi

\bibliographystyle{IEEEtran}
\bibliography{IEEEabrv,refs}

\end{document}